\documentclass[draftclsnofoot,onecolumn]{IEEEtran}
\usepackage{graphicx,epsfig,amsmath,amsbsy,amssymb,citesort,verbatim,url,algorithm,algpseudocode,soul}
\usepackage[usenames]{color}
\algnewcommand{\LineComment}[1]{\State \(\triangleright\) #1}
\algnewcommand\algorithmicswitch{\textbf{switch}}
\algnewcommand\algorithmiccase{\textbf{case}}
\algnewcommand\algorithmicdowhile{\textbf{dowhile}}
\algdef{SE}[SWITCH]{Switch}{EndSwitch}[1]{\algorithmicswitch\ #1\ \algorithmicdo}{\algorithmicend\ \algorithmicswitch}%
\algdef{SE}[CASE]{Case}{EndCase}[1]{\algorithmiccase\ #1}{\algorithmicend\ \algorithmiccase}%
\algdef{SE}[DOWHILE]{Do}{doWhile}{\algorithmicdo}[1]{\algorithmicwhile\ #1}%
\algtext*{EndCase}%
\algtext*{EndProcedure}
\algtext*{EndFor}

\newlength{\algorithmwidth}
\newtheorem{THEO}{Theorem}
\newtheorem{LEMM}{Lemma}

\newtheorem{DEFI}{Definition}

\newtheorem{CONJ}{Conjecture}
\newtheorem{COND}{Condition}

\def \real {\mathbb{R}}
\def \q{q}
\def \source{X}

\def \replevels{{\mathcal{R}}}
\def \breplevels{\mathcal{R}_F}
\def \alphabet {\mathcal{Z}}

\def \N {\mathcal{L}} 
\def \map {{{\mathcal{A}}}} 
\def \adenergy {\Psi^{H_q}_a}
\def \ss {\mathcal{S}}

\def \X {{\cal{X}}}

\def \Z {{\cal{Z}}}

\algrenewcommand{\algorithmiccomment}[1]{\hskip1em//~{\em #1}}

\begin{document}

\title{
Recovery from Linear Measurements with Complexity-Matching Universal Signal Estimation
}
\author{Junan Zhu,~\IEEEmembership{Student Member,~IEEE,}
Dror Baron,~\IEEEmembership{Senior Member,~IEEE,}\\%
and~Marco F. Duarte,~\IEEEmembership{Senior Member,~IEEE}\vspace{0mm}
\thanks{This paper was presented in part at the IEEE Workshop on Statistical Signal Processing, Gold Coast, Australia, June 2014 \cite{JZ2014SSP}, the Allerton Conference on Communications, Control, and Computing, Monticello, IL, September 2011~\cite{BaronDuarteAllerton2011}, and the Workshop on
Information Theoretic Methods in Science and Engineering, Helsinki, Finland, Aug. 2011~\cite{BaronFinland2011}.}%
\thanks{J. Zhu and D. Baron were partially supported in part by the National Science Foundation
under Grant CCF-1217749 and in part by the U.S. Army Research Office
under Grants W911NF-04-D-0003 and W911NF-14-1-0314.
M. F. Duarte was partially supported by NSF Supplemental Funding DMS-0439872 to UCLA-IPAM, PI\ R.\ Caflisch.}%
\thanks{J. Zhu and D. Baron are with the Department of Electrical and Computer Engineering, North Carolina State University, Raleigh, NC 27695. E-mail: \{jzhu9,barondror\}@ncsu.edu}%
\thanks{M. F. Duarte is with the Department of Electrical and Computer Engineering, University of Massachusetts, Amherst, MA 01003. E-mail: \quad\quad\quad  \quad\quad\quad mduarte@ecs.umass.edu}}

\maketitle \thispagestyle{empty}

\begin{abstract}

We study the compressed sensing (CS) signal estimation problem where an input signal
is measured via a linear matrix multiplication under additive noise.
While this setup usually assumes sparsity or compressibility in the input
signal during recovery, the signal structure that can be leveraged is often
not known {\em a priori}. In this paper, we consider {\em universal} CS
recovery, where the statistics of a stationary ergodic signal source are
estimated simultaneously with the signal itself.
Inspired by Kolmogorov complexity and minimum description length, we focus on a maximum
{\em a posteriori} (MAP) estimation framework that leverages universal priors to match the complexity of the source.
Our framework can also be applied to general linear inverse problems where more measurements than in CS might
be needed. We provide
theoretical results that support the algorithmic feasibility of universal MAP estimation
using a Markov chain Monte Carlo implementation, which is computationally challenging.
We incorporate some techniques to accelerate the algorithm while providing comparable
and in many cases better reconstruction quality than existing algorithms. Experimental
results show the promise of universality in CS, particularly for
low-complexity sources that do not exhibit standard sparsity or compressibility.

\end{abstract}


\section{Introduction}
\label{sec:intro}

Since many systems in science and engineering are approximately linear,
linear inverse problems have attracted great attention in the
signal processing community.
An input signal $x \in \real^N$ is recorded via a linear operator under additive noise:
\begin{equation}
\label{eq:def_y}
y = \Phi x + z,
\end{equation}
where $\Phi$ is an $M \times N$ matrix and $z \in \real^M$ denotes the noise.
The goal is to estimate $x$ from the measurements $y$ given knowledge of $\Phi$
and a model for the noise $z$. When $M \ll N$, the setup is known as compressed
sensing (CS) and the estimation problem is commonly referred to as recovery
or reconstruction; by posing a sparsity or compressibility\footnote{We use the term compressibility in this paper as defined by Cand\`{e}s et al.~\cite{CandesRUP} to refer to signals whose sparse approximation error decays sufficiently quickly.} requirement on the signal
and using this requirement as a prior during recovery, it is indeed possible to
accurately estimate $x$ from $y$~\cite{CandesRUP,DonohoCS}. On the other hand,
we might need more measurements than the signal length when the signal is dense or the noise is substantial.

Wu and Verd{\'u}~\cite{WuVerdu2012} have shown that independent and identically distributed (i.i.d.) Gaussian sensing matrices achieve the same phase-transition threshold as the optimal (potentially nonlinear) measurement operator, for any i.i.d. signals following the discrete/continuous mixture distribution $f_X(x)=p\cdot P_c(x)+(1-p)\cdot P_d(x)$, where $p$ is the probability for $x$ to take a continuous distribution $P_c(x)$ and $P_d(x)$ is an arbitrary discrete distribution. For non-i.i.d. signals, Gaussian matrices also work well~\cite{Donoho2013,Tan_CompressiveImage2014,MaZhuBaron2014submit}. Hence,
in CS the acquisition can be designed independently of the particular signal
prior through the use of randomized Gaussian matrices $\Phi$. Nevertheless, the majority of (if not all)
existing recovery algorithms require knowledge of the sparsity structure of
$x$, i.e., the choice of a {\em sparsifying transform} $W$ that renders a sparse coefficient vector $\theta=W^{-1} x$ for
the signal.

The large majority of recovery algorithms pose a sparsity prior on the signal $x$ or the coefficient vector $\theta$, e.g.,~\cite{CandesRUP,DonohoCS,GPSR2007}.
A second, separate class of Bayesian CS recovery algorithms poses a
probabilistic prior for the coefficients of $x$ in a known transform
domain~\cite{DMM2010ITW1,RanganGAMP2011ISIT,BCS2008,BCSEx2008,CSBP2010}. Given a probabilistic model, some related message passing approaches
learn the parameters of the signal model and achieve the minimum mean squared error (MMSE) in some settings; examples include EM-GM-AMP-MOS~\cite{EMGMTSP}, turboGAMP~\cite{turboGAMP}, and AMP-MixD~\cite{MTKB2014ITA}. As a third alternative,
complexity-penalized least square methods~\cite{Figueiredo2003,DonohoKolmogorovCS2006,HN05,HN11,Ramirez2011} can use arbitrary prior information on the
signal model and provide analytical guarantees, but are only computationally
efficient for specific signal models, such as the independent-entry Laplacian
model~\cite{HN05}. For example, Donoho et al.~\cite{DonohoKolmogorovCS2006} relies on Kolmogorov complexity, which cannot be computed~\cite{Cover06,LiVitanyi2008}.
As a fourth alternative, there exist algorithms that can formulate dictionaries that yield
sparse representations for the signals of interest when a large amount of training data is
available~\cite{Ramirez2011,AharoEB_KSVD,Mairal2008,Zhoul2011}.
When the signal is non-i.i.d., existing algorithms require either prior knowledge of the probabilistic model~\cite{turboGAMP} or the use of training data~\cite{Garrigues07learninghorizontal}.

In certain cases, one might not be certain about the structure or statistics of the
source prior to recovery. Uncertainty about such structure may result in a sub-optimal choice of
the sparsifying transform $W$, yielding a coefficient vector $\theta$ that requires more measurements to achieve reasonable
estimation quality; uncertainty about the statistics of the source will make it difficult to
select a prior or model for Bayesian algorithms. Thus, it would  be desirable to formulate algorithms
to estimate $x$ that are more agnostic to the particular statistics of the signal.
Therefore, we shift our focus from the standard sparsity or compressibility priors to
{\em universal} priors~\cite{LZ77,Rissanen1983,Ramirez2010}. Such concepts have been
previously leveraged in the Kolmogorov sampler universal denoising
algorithm~\cite{DonohoKolmogorov}, which minimizes Kolmogorov
complexity~\cite{Chaitin1966,Solomonoff1964,Kolmogorov1965,LiVitanyi2008,JalaliMaleki2011,JalaliMalekiRichB2014,BaronFinland2011,BaronDuarteAllerton2011}. Related approaches
based on minimum description length
(MDL)~\cite{Rissanen1978,schwarz1978estimating,Wallace1968,BRY98} minimize
the complexity of the estimated signal with respect to (w.r.t.) some class of sources.

Approaches for non-parametric sources based
on Kolmogorov complexity are not computable in practice~\cite{Cover06,LiVitanyi2008}. To address this
computational problem, we confine our attention to the class of stationary ergodic sources and
develop an algorithmic framework for {\em universal} signal estimation in CS systems that will approach the MMSE as closely as possible for the class of stationary ergodic sources. Our framework can be applied to general
linear inverse problems where more measurements might be needed.
Our framework leverages the fact that for stationary ergodic sources,
both the per-symbol empirical entropy and Kolmogorov complexity converge
asymptotically almost surely to the entropy rate of the source~\cite{Cover06}. We aim
to minimize the empirical entropy; our minimization is regularized by introducing a log
likelihood for the noise model, which is equivalent to the standard least squares under
additive white Gaussian noise. Other noise distributions are readily supported.

We make the following contributions toward our universal CS framework.
\begin{itemize}
  \item We apply a specific quantization grid to a maximum {\em a posteriori} (MAP) estimator driven by a universal prior, providing a finite-computation
universal estimation scheme; our scheme can also be applied to general linear inverse problems where more measurements might be needed.
  \item We propose a recovery algorithm based on
Markov chain Monte Carlo (MCMC)~\cite{Geman1984} to approximate this estimation procedure.
  \item We prove that for a sufficiently large number of iterations the output of our
MCMC recovery algorithm converges to the correct MAP estimate.
  \item We identify computational bottlenecks in the implementation of our MCMC estimator and show
approaches to reduce their complexity.
  \item We develop an adaptive quantization
scheme that tailors a set of reproduction levels to minimize the quantization error
within the MCMC iterations and that provides an accelerated implementation.
  \item We propose a framework that adaptively adjusts the cardinality (size) of the adaptive quantizer to match
the complexity of the input signal, in order to further reduce the quantization error and computation.
  \item We note in passing that averaging over the outputs of different runs of the same signal with the same measurements will yield
lower mean squared error (MSE) for our proposed algorithm.
\end{itemize}

This paper is organized as follows. Section~\ref{sec:setting} provides background
content. Section~\ref{sec:theory} overviews MAP estimation, quantization, and introduces universal MAP estimation.
Section~\ref{sec:MCMC} formulates an initial MCMC algorithm for universal MAP
estimation, Section~\ref{sec:adaptive} describes several improvements to this initial algorithm, and Section~\ref{sec:numerical} presents
experimental results. We conclude in Section~\ref{sec:conclusions}.
The proof of our main theoretical result appears in the appendix.

\section{Background and related work}
\label{sec:setting}
\subsection{Compressed sensing}

Consider the noisy measurement setup via a linear operator (\ref{eq:def_y}).
The input signal $x\in \real^N$ is generated by a stationary ergodic source $\source$,
and must be estimated from $y$ and $\Phi$. Note that the stationary ergodicity assumption enables us to model the potential memory in the source.
{\em The distribution $f_X$ that generates $x$ is unknown.} The matrix
$\Phi \in \real^{M \times N}$ has i.i.d. Gaussian
entries, $\Phi(m,n) \sim \mathcal{N}(0,\frac{1}{M})$.\footnote{In contrast to our analytical
and numerical results, the algorithm presented in Section~\ref{sec:MCMC} is not
dependent on a particular choice for the matrix $\Phi$.} These moments ensure that the
columns of the matrix have unit norm on average. For concrete analysis, we assume that
the noise $z\in\real^M$ is i.i.d.\ Gaussian, with mean zero and known\footnote{We assume that the noise variance is known or can be estimated~\cite{DMM2010ITW1,MTKB2014ITA}.} variance
$\sigma_Z^2$ for simplicity.

We focus on the setting where $M,N\rightarrow\infty$ and the aspect ratio is positive:
\begin{equation}
\label{eq:delta}
R \triangleq \lim_{N\rightarrow\infty} \frac{M}{N}>0.
\end{equation}
Similar settings have been discussed in the literature~\cite{Rangan2010CISS,GuoWang2008}.
When $M\ll N$, this setup is known as CS; otherwise, it is a general linear inverse problem setting.
Since $x$ is generated by an unknown source, we must search for an estimation
mechanism that is agnostic to the specific distribution $f_X$.

\subsection{Related work}
\label{subsec:Kolmogorov}

For a scalar channel with a discrete-valued signal $x$, e.g., $\Phi$ is an identity matrix and $y=x+z$,
Donoho proposed the Kolmogorov sampler (KS) for
denoising~\cite{DonohoKolmogorov},
\begin{equation}
\label{eq:x_KS}
x_{KS}  \triangleq \arg\min_w K(w)~\mbox{s.t.}~\|w-y\|^2<\tau,
\end{equation}
where $K(x)$ denotes the Kolmogorov complexity of $x$, defined as the length
of the shortest input to a Turing machine~\cite{Turing1950} that generates the
output $x$ and then halts,\footnote{For real-valued $x$, Kolmogorov complexity (KC) can be approximated using a fine quantizer. Note that the algorithm developed in this paper uses a coarse quantizer and does not rely on KC
due to the absence of a feasible method for its computation~\cite{Cover06,LiVitanyi2008} (cf.\ Section~\ref{sec:adaptive}).} and $\tau = N\sigma_Z^2$ controls for the presence
of noise. It can be shown that $K(x)$ asymptotically captures the statistics of
the stationary ergodic source $X$, and the per-symbol complexity achieves the
entropy rate $H \triangleq H(X)$, i.e., $\lim_{N\to\infty} \frac{1}{N}K(x)=H$ almost
surely~\cite[p.~154, Theorem~7.3.1]{Cover06}. Noting that universal lossless compression
algorithms~\cite{LZ77,Rissanen1983} achieve the entropy rate for any
discrete-valued finite state machine source $X$, we see that these algorithms
achieve the per-symbol Kolmogorov complexity almost surely.

Donoho et al. expanded KS to the linear CS measurement setting
$y=\Phi x$ but did not consider measurement noise~\cite{DonohoKolmogorovCS2006}.
Recent papers by Jalali and coauthors~\cite{JalaliMaleki2011,JalaliMalekiRichB2014}, which appeared
simultaneously with our work~\cite{BaronFinland2011,BaronDuarteAllerton2011},
provide an analysis of a modified KS suitable for measurements corrupted by
noise of bounded magnitude. Inspired by Donoho et al.~\cite{DonohoKolmogorovCS2006}, we
estimate $x$ from noisy measurements $y$ using the empirical entropy as a proxy
for the Kolmogorov complexity (cf.\ Section~\ref{sec:compressor}).

Separate notions of complexity-penalized least square{s} have also been shown
to be well suited for denoising and CS
recovery~\cite{Figueiredo2003,DonohoKolmogorovCS2006,Rissanen1978,schwarz1978estimating,Wallace1968,HN05,HN11,Ramirez2011}. For example,
minimum description length (MDL)~\cite{Rissanen1978,schwarz1978estimating,Wallace1968,Ramirez2011} provides
a framework composed of classes of models for which
the signal complexity can be defined sharply. In general, complexity-penalized least square
approaches can yield MDL-flavored CS recovery algorithms that are
adaptive to parametric classes of sources~\cite{DonohoKolmogorovCS2006,Figueiredo2003,HN05,HN11}. An
alternative universal denoising approach computes the universal conditional expectation of the signal~\cite{BaronFinland2011,MTKB2014ITA}.

\section{Universal MAP estimation and discretization}
\label{sec:theory}

This section briefly reviews MAP estimation and then applies it over a quantization grid, where a universal prior is used for the signal. Additionally, we provide a conjecture for the MSE achieved by our universal MAP scheme.

\subsection{Discrete MAP estimation}\label{sec:MAP}
In this subsection, we assume for exposition purposes that we know the signal distribution $f_X$.
Given the measurements $y$, the MAP estimator for $x$ has the form
\begin{equation}
x_{MAP} \triangleq \arg\max_w f_X(w) f_{Y|X}(y|w).
\label{eq:map}
\end{equation}
Because $z$ is i.i.d.\ Gaussian with mean zero and known variance $\sigma_Z^2$,
\begin{equation*}
 f_{Y|X}(y|w) = c_1 e^{-c_2 \|y-\Phi w\|^2},
\end{equation*}where
$c_1= (2\pi\sigma_Z^2)^{-M/2}$ and
$c_2= \frac{1}{2\sigma_Z^2}$
are constants, and $\|\cdot\|$ denotes the Euclidean norm.\footnote{Other noise distributions are readily supported, e.g., for i.i.d. Laplacian noise, we need to change the $\ell_2$ norm to an $\ell_1$ norm and adjust $c_1$ and $c_2$ accordingly.}
Plugging into (\ref{eq:map}) and taking log likelihoods, we obtain
$\displaystyle x_{MAP} = \arg\min_w \Psi^X(w)$,
where $\Psi^X(\cdot)$ denotes the objective function (risk)
\begin{equation*}
\Psi^X(w) \triangleq -\ln(f_X(w)) + c_2\|y-\Phi w\|^2;
\end{equation*}our ideal risk would be $\Psi^X(x_{MAP})$.

Instead of performing continuous-valued MAP estimation, we optimize for the MAP
in the discretized domain $\replevels^N$, with $\replevels$ being defined as follows.
Adapting the approach of Baron and Weissman~\cite{BaronWeissman2012},
we define the set of data-independent reproduction levels for quantizing $x$ as
\begin{equation}
\label{eq:def:replevels}
\replevels \triangleq \left\{
\ldots,-\frac{1}{\gamma},0,\frac{1}{\gamma},\ldots
\right\},
\end{equation}
where $\gamma=\lceil\ln(N)\rceil$. As $N$ increases, $\replevels$ will quantize
$x$ to a greater resolution.
These reproduction levels simplify the estimation problem from continuous to discrete.

Having discussed our reproduction levels in the set $\replevels$, we provide a technical condition on boundedness of the signal.
\begin{COND}
\label{cond:tech1}
We require that the probability density $f_X$ has bounded support, i.e., there exists $\Lambda = [x_\textrm{min},x_\textrm{max}]$ such that $f_X(x) = 0$ for $x \notin \Lambda^N$.
\end{COND}

A limitation of the data-independent reproduction level set (\ref{eq:def:replevels})
is that $\replevels$ has infinite cardinality (or size for short).
Thanks to Condition~\ref{cond:tech1}, for each value of $\gamma$ there exists
a constant $c_3>0$ such that a finite set of reproduction levels
\begin{equation}
\replevels_F \triangleq \left\{
-\frac{c_3\gamma^2}{\gamma},-\frac{c_3\gamma^2-1}{\gamma},\ldots,
\frac{c_3\gamma^2}{\gamma}
\right\}
\label{eq:def:replevels2}
\end{equation}
will quantize the range of values $\Lambda$ to the same accuracy as that of (\ref{eq:def:replevels}). We call $\replevels_F$ the {\em reproduction alphabet}, and each element in it a ({\em reproduction}) {\em level}.
This finite quantizer reduces the complexity of the estimation problem
from infinite to combinatorial. In fact,
$x_i\in [x_\textrm{min},x_\textrm{max}]$ under Condition~\ref{cond:tech1}. Therefore, for all $c_3 >0$ and sufficiently large $N$, this set of levels will cover the range $[x_\textrm{min},x_\textrm{max}]$.
The resulting reduction in complexity is due to the
structure in $\breplevels$ and independent of the particular statistics of the
source $X$.

Now that we have set up a quantization grid $(\breplevels)^N$ for $x$,
we convert the distribution $f_X$ to a probability mass function (PMF)
$\mathbb{P}_X$ over $(\breplevels)^N$. Let
$\displaystyle f_{\breplevels} \triangleq \sum_{w \in (\breplevels)^N} f_X(w)$,
and define a PMF $\mathbb{P}_X(\cdot)$ as
$\displaystyle \mathbb{P}_X(w) \triangleq \frac{f_X(w)}{f_{\breplevels}}$.
Then
\begin{equation*}
x_{MAP}(\breplevels) \triangleq \arg\min_{w \in (\breplevels)^N} \left(-\ln(  \mathbb{P}_X(w)  ) + c_2 \|y-\Phi w\|^2\right)
\end{equation*}gives the MAP estimate of $x$ over $(\breplevels)^N$.
Note that we use the PMF formulation above, instead of the more common bin integration formulation, in order to simplify our presentation and analysis. Luckily, as $N$ increases, $\mathbb{P}_X$ will approximate $f_X$ more closely under (\ref{eq:def:replevels2}).

\subsection{Universal MAP estimation}
\label{sec:univtheory}
We now describe a
universal estimator for CS over a quantized grid. Consider a prior
$\mathbb{P}_U$ that might involve Kolmogorov
complexity~\cite{Chaitin1966,Solomonoff1964,Kolmogorov1965},
e.g., $\mathbb{P}_U(w)=2^{-K(w)}$, or MDL complexity w.r.t.\ some class of
parametric sources~\cite{Rissanen1978,schwarz1978estimating,Wallace1968}.
We call $\mathbb{P}_U$ a {\em universal prior} if it has the fortuitous property that for every stationary ergodic
source $X$ and fixed $\epsilon > 0$,  there exists some minimum $N_0(X,\epsilon)$ such that
\begin{equation*}
 -\frac{\ln(\mathbb{P}_U(w))}{N} < -\frac{\ln(\mathbb{P}_X(w))}{N} + \epsilon
\end{equation*}for all $w\in(\breplevels)^N$ and $N > N_0(X,\epsilon)$~\cite{LZ77,Rissanen1983}.
We optimize over an objective function that incorporates $\mathbb{P}_U$ and the presence
of additive white Gaussian noise in the measurements:
\begin{equation}
\Psi^U(w) \triangleq -\ln(\mathbb{P}_U(w)) + c_2 \|y-\Phi w\|^2,
\label{eq:psidef}
\end{equation}
resulting in\footnote{This formulation of $x_{MAP}^U$  corresponds to a Lagrangian relaxation of the approach studied
in~\cite{JalaliMaleki2011,JalaliMalekiRichB2014}.}
$\displaystyle
\label{eq:universal_x}
x_{MAP}^U \triangleq \arg \min_{w \in (\breplevels)^N} \Psi^U(w)$.
Our universal MAP estimator does not require $M\ll N$, and $x_{MAP}^U$ can be used in general linear inverse problems.

\subsection{Conjectured MSE performance}\label{sec:conjecture}
Donoho~\cite{DonohoKolmogorov} showed for the scalar
channel $y=x+z$ that: ($i$) the Kolmogorov sampler $x_{KS}$ (\ref{eq:x_KS})
is drawn from the posterior distribution $\mathbb{P}_{X|Y}(x|y)$; and ($ii$) the
MSE of this estimate $E_{X,Z,\Phi}[\|y-x_{KS}\|^2]$
is no greater than twice the MMSE.
Based on this result, which requires a large reproduction alphabet, we now present a conjecture on the quality of the estimation $x^U_{MAP}$. Our conjecture is based on observing that
({\em i}) in the setting~\eqref{eq:def_y}, Kolmogorov sampling achieves optimal rate--distortion performance;
({\em ii}) the Bayesian posterior distribution is the solution to the rate-distortion problem; and
({\em iii}) sampling from the Bayesian posterior yields a squared error that is no greater
than twice the MMSE. Hence, $x^U_{MAP}$ behaves as if we sample
from the Bayesian posterior distribution and yields no greater
than twice the MMSE; some experimental evidence to assess this conjecture is presented in Figs.~\ref{fig:Ber} and~\ref{fig:sparseL}.

\begin{CONJ}
\label{conj:double_MMSE}
Assume that $\Phi\in\real^{M\times N}$ is an i.i.d.\ Gaussian measurement matrix
where each entry has mean zero and variance $1/M$. Suppose that
Condition~\ref{cond:tech1} holds, the aspect ratio
$R>0$ in~(\ref{eq:delta}), and the noise $z\in\real^M$ is i.i.d.\ zero-mean
Gaussian with finite variance.
Then for all $\epsilon>0$, the mean squared error of the
universal MAP estimator $x^U_{MAP}$ satisfies
\begin{equation*}
\frac{E_{X,Z,\Phi}\left[\|x-x^U_{MAP}\|^2\right]}{N} < \frac{2 E_{X,Z,\Phi}\left[\|x-E_{X}[x|y,\Phi]\|^2\right]}{N} +\epsilon
\end{equation*}
for sufficiently large $N$.
\end{CONJ}

\section{Fixed reproduction alphabet algorithm}
\label{sec:MCMC}

Although the results of the previous section are theoretically appealing, a brute force
optimization of $x_{MAP}^U$ is computationally intractable. Instead, we propose an
algorithmic approach based on MCMC
methods~\cite{Geman1984}. Our approach is reminiscent of the framework for lossy data
compression in~\cite{Jalali2008,Jalali2012,BaronWeissman2012,Yang1997}.

\subsection{Universal compressor}\label{sec:compressor}

We propose a universal lossless compression formulation following
the conventions of Weissman and coauthors~\cite{Jalali2008,Jalali2012,BaronWeissman2012}.
We refer to the estimate as $w$ in our algorithm.
Our goal is to characterize $-\ln(\mathbb{P}_U(w))$, cf.~(\ref{eq:psidef}).
Although we are inspired by the Kolmogorov sampler approach~\cite{DonohoKolmogorov}, KC cannot be computed~\cite{Cover06,LiVitanyi2008}, and we instead use empirical entropy. For stationary ergodic sources, the empirical entropy
converges to the per-symbol entropy rate almost surely~\cite{Cover06}.

To define the empirical entropy, we first define the empirical symbol
counts:
\begin{equation}
n_\q(w,\alpha)[\beta] \triangleq | \{ i \in [\q+1,N]: w_{i-\q}^{i-1}=\alpha, w_i=\beta \} |,
\label{eq:nq}
\end{equation}
where $\q$ is the context depth~\cite{Rissanen1983,Willems1995CTW},
$\beta \in \replevels_F$, $\alpha\in(\replevels_F)^\q$, $w_i$ is the $i^{th}$ symbol of $w$,
and $w_i^j$ is the string comprising symbols $i$ through $j$ within $w$.
We now define the order $\q$ conditional empirical probability for the context
$\alpha$ as
\begin{equation}
\label{eq:def:Pcond}
\mathbb{P}_\q(w,\alpha)[\beta] \triangleq
 \frac{  n_\q(w,\alpha)[\beta] } { \sum_{\beta' \in \replevels_F} n_\q(w,\alpha)[\beta'] },
\end{equation}
and the order $\q$ conditional empirical entropy,
\begin{align}
H_\q(w) \triangleq -\frac{1}{N} \sum_{\alpha \in (\replevels_F)^\q,\beta \in \replevels_F} n_\q(w,\alpha)[\beta]
\log_2\left( \mathbb{P}_\q(w,\alpha)[\beta] \right),
\label{eq:def:H}
\end{align}
where the sum is only over non-zero counts and probabilities.

Allowing the context depth $\q \triangleq \q_N=o(\log(N))$ to grow slowly with $N$,
various universal compression algorithms can achieve the empirical entropy
$H_{\q}(\cdot)$ asymptotically~\cite{Rissanen1983,Willems1995CTW,LZ77}.
On the other hand, no compressor can outperform the entropy rate. Additionally,
for large $N$, the empirical symbol counts with context depth $\q$ provide a
sufficiently precise characterization of the source statistics. Therefore, $H_\q$
provides a concise approximation to the per-symbol coding length of a universal
compressor.

\subsection{Markov chain Monte Carlo}\label{sec:B-MCMC}

Having approximated the coding length, we now describe how to
optimize our objective function.
We define the energy $\Psi^{H_q}(w)$
in an analogous manner to $\Psi^U(w)$ (\ref{eq:psidef}),
using $H_\q(w)$ as our universal coding length:
\begin{equation}
\label{eq:MCMC_energy}
\Psi^{H_q}(w) \triangleq NH_\q(w) + c_4 \|y - \Phi w\|^2,
\end{equation}
where $c_4=c_2 \log_2(e) $.
The minimization of this energy is analogous to minimizing $\Psi^U(w)$.

Ideally, our goal is to compute the globally minimum energy solution
$\displaystyle x_{MAP}^{H_q} \triangleq \arg\min_{w \in (\replevels_F)^N} \Psi^{H_q}(w)$.
We use a stochastic MCMC
relaxation~\cite{Geman1984} to achieve the globally minimum solution
in the limit of infinite computation.
To assist the reader in appreciating how MCMC is used to compute $x_{MAP}^{H_q}$, we include pseudocode for our approach in Algorithm~\ref{alg:MCMC}. The algorithm, called basic MCMC (B-MCMC), will be used as a building block for our latter Algorithms~\ref{alg:MCMCAL} and~3 in Section~\ref{sec:adaptive}.
The initial estimate $w$ is obtained by quantizing the {\em initial point} $x^*\in\real^N$ to $(\replevels_F)^N$. The initial point $x^*$ could be the output of any signal reconstruction algorithm, and because $x^*$ is a preliminary estimate of the signal that does not require high fidelity, we let $x^*=\Phi^T y$ for simplicity, where $(\cdot)^T$ denotes transpose.
We refer to the processing of a single entry of $w$ as an iteration and group the
processing of all entries of $w$, randomly permuted, into
super-iterations.

The Boltzmann PMF is defined as
\begin{equation}
\label{eq:def_Boltzmann}
\mathbb{P}_s(w) \triangleq \frac{1}{\zeta_s} \exp(-s \Psi^{H_q}(w)),
\end{equation}
where $s>0$ is inversely related to the temperature in simulated annealing and
$\zeta_s$ is a normalization constant.
MCMC samples from the Boltzmann PMF (\ref{eq:def_Boltzmann}) using a
{\em Gibbs sampler}: in each iteration, a single element $w_n$ is generated
while the rest of $w$, $w^{\backslash n} \triangleq \{ w_i:\ n \neq i\}$,
remains unchanged. We denote by $w_1^{n-1} \beta w_{n+1}^N$ the
concatenation of the initial portion of the output vector $w_1^{n-1}$, the symbol
$\beta \in \replevels_F$, and the latter portion of the output $w_{n+1}^N$. The
Gibbs sampler updates $w_n$ by resampling from the PMF:
\begin{eqnarray}
& &\! \! \! \! \! \mathbb{P}_s(w_n=a|w^{\backslash n}) \label{eqn:Gibbs}\\
&=&\! \! \! \! \!  \frac{ \exp\left(-s\Psi^{H_q}(w_1^{n-1}aw_{n+1}^N) \right) }
{ \sum_{b \in \breplevels} \exp\left( -s\Psi^{H_q}(w_1^{n-1}bw_{n+1}^N) \right) } \nonumber\\
&=&\! \! \! \! \!  \frac{1}{\sum_{b \in \breplevels} \exp\left( -s\left[ N\Delta H_\q(w,n,b,a) + c_4
\Delta d(w,n,b,a) \right] \right) }\nonumber,
\end{eqnarray}
where
\begin{eqnarray*}
\Delta H_\q(w,n,b,a)  \triangleq
 H_\q(w_1^{n-1}bw_{n+1}^N)  -  H_\q(w_1^{n-1}aw_{n+1}^N)
\end{eqnarray*}
is the change in empirical entropy $H_\q(w)$ (\ref{eq:def:H}) when $w_n=a$
is replaced by $b$, and
\begin{equation}
\begin{split}
\Delta d(w,n,b,a) &\triangleq  \|y-\Phi(w_1^{n-1}bw_{n+1}^N)\|^2  \\
&- \|y-\Phi(w_1^{n-1}aw_{n+1}^N)\|^2 \label{eq:Deltad}
\end{split}
\end{equation}
is the change in $\|y-\Phi w\|^2$ when $w_n=a$ is replaced by $b$. The maximum change in the energy within an iteration of Algorithm~\ref{alg:MCMC} is then bounded by
\begin{equation}
\begin{split}
\Delta_q = \max_{1\le n \le N} \max_{w \in (\breplevels)^N} \max_{a,b \in \replevels_F}& |N\Delta H_\q(w,n,b,a)\\
&+c_4\Delta d(w,n,b,a)|.
\label{eq:Deltaq}
\end{split}
\end{equation}
Note that $x$ is assumed bounded (cf.\ Condition~\ref{cond:tech1}) so that (\ref{eq:Deltad}--\ref{eq:Deltaq}) are bounded as well.

In MCMC, the space $w\in(\replevels_F)^N$
is analogous to a statistical mechanical system,
and at low temperatures the system tends toward low energies. Therefore, during the execution of the algorithm, we set a sequence of decreasing temperatures that takes into account the maximum change given in (\ref{eq:Deltaq}):
\begin{align}
s_t \triangleq \ln(t+r_0)/(cN\Delta_q)~\textrm{for some}~c>1,
\label{eq:st}
\end{align}where $r_0$ is a temperature offset. At low temperatures, i.e., large $s_t$, a small difference in energy
$\Psi^{H_q}(w)$ drives a big difference in probability, cf.~(\ref{eq:def_Boltzmann}). Therefore, we begin at a high
temperature where the Gibbs sampler can freely move around $(\replevels_F)^N$. As the
temperature is reduced, the PMF becomes more sensitive to changes in energy
(\ref{eq:def_Boltzmann}), and the trend toward $w$ with lower energy grows stronger.
In each iteration, the Gibbs sampler modifies $w_n$ in a
random manner that resembles heat bath concepts in statistical mechanics. Although
MCMC could sink into a local minimum, Geman and Geman~\cite{Geman1984} proved that if we decrease the temperature according to (\ref{eq:st}), then the randomness of Gibbs sampling will eventually drive MCMC out of the locally
minimum energy and it will converge to the globally optimal energy w.r.t.\ $x_{MAP}^U$. Note that Geman and Geman proved that MCMC will converge, although the proof states that it will take infinitely long to do so. In order to help B-MCMC approach the global minimum with reasonable runtime, we will refine B-MCMC in Section~\ref{sec:adaptive}.

\begin{algorithm}[!t]
\caption{Basic MCMC for universal CS -- Fixed alphabet} \label{alg:MCMC}
\begin{algorithmic}[1]
\State {\bf Inputs}: Initial estimate $w$, reproduction alphabet $\replevels_F$, noise variance $\sigma_Z^2$, number of super--iterations $r$, temperature constant $c>1$, and context depth $q$
\State Compute $n_\q(w,\alpha)[\beta],~\forall~\alpha \in (\replevels_F)^\q$, $\beta \in \replevels_F$
\For{$t=1$ to $r$} \Comment{super-iteration}
\State $s \leftarrow \ln(t)/(cN\Delta_q)$ \Comment{$s=s_t$, cf.~(\ref{eq:st})}
\State Draw permutation $\{1,\ldots,N\}$ at random
\For{$t'=1$ to $N$} \Comment{iteration}
\State Let~$n$ be component $t'$ in permutation
\For{all $\beta$ in $\replevels_F$} \Comment{possible new $w_n$}
\State Compute $\Delta H_\q(w,n,\beta,w_n)$ \label{sudo:fixedR_DeltaH}
\State Compute $\Delta d(w,n,\beta,w_n)$  \label{sudo:fixedR_Deltad}
\State Compute $\mathbb{P}_s(w_n=\beta|w^{\backslash n})$   \label{sudo:fixedR:fs}
\EndFor
\State Generate $w_n$ using $\mathbb{P}_s(\cdot|w^{\backslash n})$ \Comment{Gibbs}
\State Update $n_\q(w,\alpha)[\beta],~\forall~\alpha \in (\replevels_F)^\q$, $\beta \in \replevels_F$ \label{sudo:fixedR:update}
\EndFor
\EndFor
\State {\bf Output:}\ Return approximation $w$ of $x^U_{MAP}$
\end{algorithmic}
\end{algorithm}

The following theorem is proven in~\ref{ap:th:conv}, following the framework established by Jalali and Weissman~\cite{Jalali2008,Jalali2012}.
\begin{THEO}
Let $X$ be a stationary ergodic source that obeys Condition~\ref{cond:tech1}. Then the outcome $w^r$ of
Algorithm~\ref{alg:MCMC}
in the limit of an infinite number of super-iterations $r$ obeys
\begin{equation*}
\lim_{r \to \infty} \Psi^{H_q}(w^r) = \min_{\widetilde{w} \in (\replevels_F)^N} \Psi^{H_q}(\widetilde{w}) = \Psi^{H_q}\left(x_{MAP}^{H_q}\right).
\end{equation*}
\label{th:conv}
\end{THEO}

Theorem~\ref{th:conv} shows that Algorithm~\ref{alg:MCMC} matches the
best-possible performance of the universal MAP estimator as measured by the objective function $\Psi^{H_q}$, which should yield an MSE that is twice the MMSE (cf.\ Conjecture~\ref{conj:double_MMSE}). We want to remind the reader that Theorem~\ref{th:conv} is based on the stationarity and ergodicity of the source, which could have memory.
To gain some insight about the convergence process of MCMC, we focus on a fixed arbitrary sub-optimal sequence $w\in(\replevels_F)^N$. Suppose that at super-iteration $t$ the energy for the algorithm's output $\Psi^{H_q}(w)$ has converged to the steady state (see~\ref{ap:th:conv} for details on convergence). We can then focus on the probability ratio
$\displaystyle\rho_t=\mathbb{P}_{s_t}(w)/\mathbb{P}_{s_t}(x^{H_q}_{MAP})$;
$\rho_t<1$ because $x^{H_q}_{MAP}$ is the global minimum and has the largest Boltzmann probability over all $w\in(\replevels_F)^N$, whereas $w$ is sub-optimal. We then consider the same sequence $w$ at super-iteration $t^2$; the inverse temperature is $2s_t$ and the corresponding ratio at super-iteration $t^2$ is (cf.~(\ref{eq:def_Boltzmann}))
\begin{equation*}
\frac{\mathbb{P}_{2s_t}(w)}{\mathbb{P}_{2s_t}(x^{H_q}_{MAP})} = \frac{\exp(-2s_t
\Psi^{H_q}(w))}{\exp(-2s_t \Psi^{H_q}(x^{H_q}_{MAP}))} =
\left(  \frac{ \mathbb{P}_{s_t}(w)}{\mathbb{P}_{s_t}(x^{H_q}_{MAP})}\right)^2.
\end{equation*}
That is, between super-iterations $t$ and $t^2$ the probability ratio $\rho_t$ is also squared, and the Gibbs sampler is less likely to generate samples whose energy differs significantly from the minimum energy w.r.t.\
$x^{H_q}_{MAP}$. We infer from this argument that the probability concentration of our algorithm around the globally optimal energy w.r.t.\ $x^{H_q}_{MAP}$ is linear in the number of super-iterations.

\subsection{Computational challenges}\label{sec:B-MCMC_complexity}

Studying the pseudocode of Algorithm~\ref{alg:MCMC}, we recognize that
Lines~\ref{sudo:fixedR_DeltaH}--\ref{sudo:fixedR:fs} must be implemented efficiently,
as they run $rN|\replevels_F|$ times. Lines~\ref{sudo:fixedR_DeltaH}
and~\ref{sudo:fixedR_Deltad} are especially challenging.

For Line~\ref{sudo:fixedR_DeltaH}, a naive update of $H_q(w)$ has
complexity $O(|\replevels_F|^{\q+1})$, cf.~(\ref{eq:def:H}). To address this problem,
Jalali and Weissman~\cite{Jalali2008,Jalali2012} recompute the empirical conditional
entropy in $O(\q|\replevels_F|)$ time only for the $O(\q)$ contexts whose corresponding
counts are modified~\cite{Jalali2008,Jalali2012}. The same approach can be used in
Line~\ref{sudo:fixedR:update}, again reducing computation from $O(|\replevels_F|^{\q+1})$ to $O(\q|\replevels_F|)$.
Some straightforward algebra allows us to convert Line~\ref{sudo:fixedR_Deltad} to a form that requires aggregate runtime of $O(Nr(M+|\replevels_F|))$.
Combined with the
computation for Line~\ref{sudo:fixedR_DeltaH}, and since $M \gg \q |\replevels_F|^2$ (because $|\replevels_F|=\gamma^2, \gamma=\lceil\ln(N)\rceil, q=o(\log(N))$, and $M=O(N)$)
in practice, the entire runtime of our algorithm is $O(rMN)$.

The practical value of Algorithm~\ref{alg:MCMC} may be reduced due to its high
computational cost, dictated by the number of super-iterations $r$ required for convergence
to $x^{H_q}_{MAP}$ and the large size of the reproduction alphabet. Nonetheless, Algorithm~\ref{alg:MCMC} provides a starting point toward further performance gains of more practical algorithms for computing $x^{H_q}_{MAP}$, which are presented in Section~\ref{sec:adaptive}. Furthermore, our experiments in Section~\ref{sec:numerical} will show that the performance of the algorithm of Section~\ref{sec:adaptive} is comparable to and in many cases better than existing algorithms.

\section{Adaptive reproduction alphabet}
\label{sec:adaptive}
While Algorithm~\ref{alg:MCMC} is a first step toward universal signal estimation in CS, $N$ must be large enough to ensure that $\replevels_F$
quantizes a broad enough range of values of $\real$ finely enough to represent the
estimate $x^{H_q}_{MAP}$ well. For large $N$, the estimation performance using the
reproduction alphabet~(\ref{eq:def:replevels2}) could suffer from high computational complexity.
On the other hand, for small $N$ the number of reproduction levels employed is insufficient to obtain acceptable performance. Nevertheless, using an excessive number of levels will slow down the convergence. Therefore, in this section, we explore techniques that tailor the reproduction alphabet adaptively to the signal being observed.

\subsection{Adaptivity in reproduction levels}\label{sec:L-MCMC}
To estimate better with finite $N$, we utilize reproduction levels that are
{\em adaptive} instead of the fixed levels in $\replevels_F$. To do so, instead of
$w \in (\replevels_F)^N$, we optimize over a sequence $u \in \Z^N$, where $|\Z| < |\breplevels|$ and $|\cdot|$ denotes the size.
The new reproduction alphabet $\Z$ does
not directly correspond to real numbers. Instead, there is an adaptive mapping
$\map: \Z \rightarrow \real$, and the reproduction levels are $\map(\Z)$. Therefore, we call $\Z$ the {\em adaptive} reproduction alphabet. Since the mapping $\map$ is one-to-one, we also refer to $\Z$ as reproduction levels.
Considering the energy function (\ref{eq:MCMC_energy}),
we now compute the empirical symbol counts $n_\q(u,\alpha)[\beta]$, order $\q$
conditional empirical probabilities $\mathbb{P}_\q(u,\alpha)[\beta]$, and order $\q$
conditional empirical entropy $H_\q(u)$ using $u \in \Z^N$, $\alpha \in \Z^\q$, and
$\beta \in \Z$, cf.~(\ref{eq:nq}), (\ref{eq:def:Pcond}), and~(\ref{eq:def:H}). Similarly, we
use $\|y - \Phi \map(u)\|^2$ instead of $\|y - \Phi w\|^2$, where $\map(u)$ is the
straightforward vector extension of $\map$. These modifications yield an adaptive
energy function $\displaystyle \adenergy(u) \triangleq NH_\q(u) + c_4  \|y - \Phi \map(u)\|^2$.

We choose $\map_{opt}$ to optimize for minimum
squared error,
\begin{align*}
\map_{opt} &\triangleq \arg \min_{\map}\|y-\Phi \map(u)\|^2 \\
&= \arg \min_{\map}\left[\sum_{m=1}^M(y_m-[\Phi \map(u)]_m)^2\right],
\end{align*}
where $[\Phi \map(u)]_m$ denotes the $m^{th}$ entry of the vector $\Phi \map(u)$.
The optimal mapping depends entirely on $y$, $\Phi$, and $u$. From a coding
perspective, describing $\map_{opt}(u)$ requires $H_q(u)$ bits for $u$ and
$|\Z| b\log \log(N)$ bits for $\map_{opt}$ to match the resolution of the non-adaptive $\replevels_F$, with $b > 1$ an arbitrary constant~\cite{BaronWeissman2012}. The resulting coding length
defines our universal prior.

\textbf{Optimization of reproduction levels: }
We now describe the optimization procedure for $\map_{opt}$, which must be
computationally efficient. Write
\begin{align*}
\Upsilon(\map) \triangleq \|y-\Phi \map(u)\|^2 = \sum_{m=1}^M\left(y_m-\sum_{n=1}^N \Phi_{mn}\map(u_n)\right)^2,
\end{align*}
where $\Phi_{mn}$ is the entry of $\Phi$ at row $m$ and column $n$.
For $\Upsilon(\map)$ to be minimum, we need zero-valued derivatives in~(\ref{eq:derzero}), where $1_{\{A\}}$ is the indicator function for event $A$.
\begin{figure*}
\vspace*{-5mm}
\begin{equation}
\frac{d\Upsilon(\map)}{d\map(\beta)} = -2 \sum_{m=1}^M \left(y_m - \sum_{n=1}^N \Phi_{mn} \map(u_n) \right)
\left( \sum_{n=1}^N \Phi_{mn} 1_{\{u_n=\beta\}} \right) = 0,~\forall~\beta \in \Z \label{eq:derzero}
\end{equation}
\end{figure*}
Define the location sets
$\displaystyle \N_\beta \triangleq \{ n: 1\le n \le N, u_n = \beta\}$
for each $\beta \in \Z$, and rewrite the derivatives of $\Upsilon(\map)$,
\begin{align}
\frac{d\Upsilon(\map)}{d\map(\beta)} =
-2 \sum_{m=1}^M \left( y_m - \sum_{\lambda \in \Z} \sum_{n \in \N_{\lambda}} \Phi_{mn} \map(\lambda) \right)
\left( \sum_{n\in \N_\beta} \Phi_{mn} \right).
\label{eq:derivative2}
\end{align}
Let the per-character sum column values be
\begin{equation}
\label{eq:mu_def}
\mu_{m\beta} \triangleq \sum_{n \in \N_{\beta}} \Phi_{mn},
\end{equation}
for each $m\in\{1,\ldots,M\}$ and $\beta\in\Z$.
We desire the derivatives to be zero, cf.~(\ref{eq:derivative2}):
\begin{align*}
0 = \sum_{m=1}^M \left(y_m-\sum_{\lambda \in \Z}\map(\lambda) \mu_{m\lambda}\right) \mu_{m\beta}.
\end{align*}
Thus, the system of equations must be satisfied,
\begin{align}\label{eq:system_equation}
\sum_{m=1}^M y_m \mu_{m\beta}
=
\sum_{m=1}^M \left(\sum_{\lambda \in \Z} \map(\lambda) \mu_{m\lambda}\right) \mu_{m\beta}
\end{align}
for each $\beta \in \Z$. Consider now the right hand side,
\begin{align*}
\sum_{m=1}^M \left(\sum_{\lambda \in \alphabet} \map(\lambda) \mu_{m\lambda}\right)
\mu_{m\beta}
= \sum_{\lambda \in \alphabet} \map(\lambda) \sum_{m=1}^M \mu_{m\lambda} \mu_{m\beta},
\end{align*}
for each $\beta \in \Z$. The system of equations can be described in matrix form in (\ref{eq:matrix_map_opt}).
\begin{figure*}
\vspace*{-5mm}
\begin{align}
\label{eq:matrix_map_opt}
\overbrace{
\left[\begin{array}{ccc}
\sum_{m=1}^M \mu_{m\beta_1} \mu_{m\beta_1} & \ldots & \sum_{m=1}^M \mu_{m\beta_{|\alphabet|}} \mu_{m\beta_1}\\
\vdots & \ddots & \vdots\\
\sum_{m=1}^M \mu_{m\beta_1} \mu_{m\beta_{|\alphabet|}} & \ldots & \sum_{m=1}^M \mu_{m\beta_{|\alphabet|}} \mu_{m\beta_{|\alphabet|}}\\
\end{array}
\right]
}^{\Omega}
\overbrace{
\left[\begin{array}{c}
\map(\beta_1)\\
\vdots\\
\map(\beta_{|\alphabet|})
\end{array}\right]}
^{\map(\Z)}=
\overbrace{
\left[\begin{array}{c}
\sum_{m=1}^M y_m \mu_{m\beta_1}\\
\vdots\\
\sum_{m=1}^M y_m \mu_{m\beta_{|\alphabet|}}
\end{array}
\right]}
^{\Theta}
\end{align}
\hrulefill
\end{figure*}
Note that by writing $\mu$ as a matrix with entries indexed by row $m$ and
column $\beta$ given by (\ref{eq:mu_def}), we can write $\Omega$ as a Gram
matrix, $\Omega=\mu^T \mu$, and we also have $\Theta = \mu^T y$, cf.~(\ref{eq:system_equation}).
The optimal $\map$ can be computed as a $|\Z|\times 1$ vector
$\displaystyle \map_{opt} = \Omega^{-1} \Theta = (\mu^T\mu)^{-1}\mu^Ty$ if
$\Omega \in \mathbb{R}^{|\Z|\times |\Z|}$ is invertible. We note in passing that numerical
stability can be improved by regularizing $\Omega$. Note also that
\begin{equation}
\label{eqn:compute_ell2}
\|y-\Phi \map(u)\|^2  =
\sum_{m=1}^M \left(
y_m - \sum_{\beta \in \Z} \mu_{m\beta} \map_{opt}(\beta)
\right)^2,
\end{equation}
which can be computed in $O(M|\Z|)$ time instead of $O(MN)$.

\textbf{Computational complexity: }
Pseudocode for level-adaptive MCMC (L-MCMC) appears in Algorithm~\ref{alg:MCMCAL}, which resembles Algorithm~\ref{alg:MCMC}.
The initial mapping $\map$ is inherited from a quantization of the initial point $x^*$, $r_0=0$ ($r_0$ takes different values in Section~\ref{sec:adaptive_size}), and other minor differences between B-MCMC and L-MCMC appear in lines marked by asterisks.

We discuss computational requirements for each line of the pseudocode that is run
within the inner loop.
\begin{itemize}
\item
Line~\ref{Algo2:DeltaH} can be computed in $O(q|\Z|)$ time (see discussion of Line~\ref{sudo:fixedR_DeltaH} of B-MCMC in Section~\ref{sec:B-MCMC_complexity}).
\item
Line~\ref{Algo2:mu} updates $\mu_{m\beta}$ for $m=1,...,M$ in $O(M)$ time.
\item
Line~\ref{Algo2:Omega}
updates $\Omega$. Because we only need to update
$O(1)$ columns and $O(1)$ rows, each such column and
row contains $O(|\Z|)$ entries, and each entry is a sum over $O(M)$
terms, we need $O(M|\Z|)$ time.
\item
Line~\ref{Algo2:map_opt}
requires inverting $\Omega$ in $O(|\Z|^3)$ time.
\item
Line~\ref{Algo2:ell2}
requires $O(M|\Z|)$ time, cf.~(\ref{eqn:compute_ell2}).
\item
 Line~\ref{Algo2:distribution} requires $O(|\Z|)$ time.
\end{itemize}
In practice we typically have $M \gg |\Z|^2$,
and so the aggregate complexity is $O(rMN|\Z|)$, which is
greater than the computational complexity of Algorithm~\ref{alg:MCMC} by a factor of $O(|\Z|)$.

\begin{algorithm}[!t]
\caption{Level-adaptive MCMC} \label{alg:MCMCAL}
\begin{algorithmic}[1]
\State *{\bf Inputs}: Initial mapping $\map$, sequence $u$, adaptive alphabet $\Z$, noise variance $\sigma_Z^2$, number of super-iterations $r$, temperature constant $c>1$, context depth $q$, and temperature offset $r_0$
\State Compute $n_\q(u,\alpha)[\beta],~\forall~\alpha \in \Z^\q$, $\beta \in \Z$
\State *Initialize $\Omega$\label{Algo2:init_omega}
\For{$t=1$ to $r$} \Comment{super-iteration}
\State $s \leftarrow \ln(t+r_0)/(cN\Delta_q)$ \Comment{$s=s_t$, cf.~(\ref{eq:st})}
\State Draw permutation $\{1,\ldots,N\}$ at random
\For{$t'=1$ to $N$} \Comment{iteration}
\State Let~$n$ be component $t'$ in permutation
\For{all $\beta$ in $\Z$} \Comment{possible new $u_n$}
\State Compute $\Delta H_\q(u,n,\beta,u_n)$ \label{Algo2:DeltaH}
\State *Compute $\mu_{m\beta},\forall\ m \in\{1,\ldots,M\}$ \label{Algo2:mu}
\State *Update $\Omega$ \label{Algo2:Omega} \Comment{$O(1)$ rows and columns}
\State *Compute $\map_{opt}$ \Comment{invert $\Omega$} \label{Algo2:map_opt}
\State Compute $\|y-\Phi\map(u_1^{n-1}\beta u_{n+1}^N)\|^2$ \label{Algo2:ell2}
\State Compute $\mathbb{P}_s(u_n=\beta|u^{\backslash n})$ \label{Algo2:distribution}
\EndFor
\State *$\widetilde{u}_n \leftarrow u_n$  \Comment{save previous value}\label{Algo2:save_sequence}
\State Generate $u_n$ using $\mathbb{P}_s(\cdot|u^{\backslash n})$ \Comment{Gibbs}
\State Update $n_\q(\cdot)[\cdot]$ at $O(\q)$ relevant locations
\State *Update $\mu_{m\beta},~\forall~m$, $\beta\in \{u_n,\widetilde{u}_n\}$\label{Algo2:update_mu}
\State *Update $\Omega$ \Comment{$O(1)$ rows and columns}\label{Algo2:update_omega}
\EndFor
\EndFor
\State *{\bf Outputs}: Return approximation $\map(u)$ of $x^U_{MAP},~\Z$, and temperature offset $r_0+r$
\end{algorithmic}
\end{algorithm}

\begin{figure*}[t]
\begin{center}
\includegraphics[width=180mm]{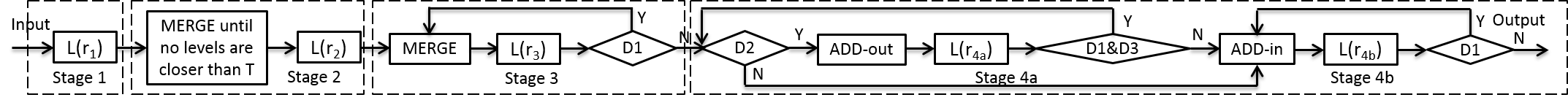}
\end{center}
\vspace*{-5mm}
\caption{{\small\sl Flowchart of Algorithm 3 (size- and level-adaptive MCMC).
L($r$) denotes running L-MCMC for $r$ super-iterations. The parameters $r_1$,$r_2$,$r_3$,$r_{4a}$, and $r_{4b}$ are the number of super-iterations used in Stages~1 through~4, respectively. Criteria $D1-D3$ are described in the text.}
\label{fig:SLA-MCMC}}
\vspace*{-5mm}
\end{figure*}

\subsection{Adaptivity in reproduction alphabet size}\label{sec:adaptive_size}

While Algorithm~\ref{alg:MCMCAL} adaptively maps $u$ to $\mathbb{R}^N$, the signal estimation quality heavily depends on $|\Z|$. Denote the true alphabet of the signal by $\X,~x\in {\X}^N$; if the signal is continuous-valued, then $|\X|$ is infinite. Ideally we want to
employ as many levels as the runtime allows for continuous-valued signals,
whereas for discrete-valued signals we want $|\Z|=|\X|$.
Inspired by this observation, we propose to
begin with some initial $|\Z|$, and then adaptively adjust $|\Z|$ hoping to match $|\X|$. Hence, we propose the size- and level-adaptive MCMC algorithm (Algorithm~3), which invokes L-MCMC (Algorithm~\ref{alg:MCMCAL}) several times.

\textbf{Three basic procedures: }
In order to describe the size- and level-adaptive MCMC (SLA-MCMC) algorithm in detail, we introduce three alphabet adaptation procedures as follows.
\begin{itemize}
  \item {\em MERGE}: First, find the closest adjacent levels $\beta_1,\beta_2\in \Z$. Create a new level $\beta_3$ and add it to $\Z$. Let $\map(\beta_3)=(\map(\beta_1)+\map(\beta_2))/2$.
  Replace $u_i$ by $\beta_3$ whenever $u_i\in\{\beta_1,\beta_2\}$. Next, remove $\beta_1$ and $\beta_2$ from $\Z$.
\item {\em ADD-out}: Define the range $R_{\map}=[\min\map(\Z),$ $\max\map(\Z)]$, and $\mathcal{I}_{R_{\map}}=\max\map(\Z)-\min\map(\Z)$. Add a {\em lower } level $\beta_3$ and/or {\em upper level} $\beta_4$ to $\Z$ with
    \begin{eqnarray*}
    \map(\beta_3)=\min\map(\Z)-\frac{\mathcal{I}_{R_{\map}}}{|\Z|-1},\\ \map(\beta_4)=\max\map(\Z)+\frac{\mathcal{I}_{R_{\map}}}{|\Z|-1}.
    \end{eqnarray*}Note that $|\{u_i: u_i=\beta_3 \mbox{ or } \beta_4, i=1,...,N\}|=0$, i.e., the new levels are empty.
\item {\em ADD-in}: First, find the most distant adjacent levels, $\beta_1$ and $\beta_2$. Then, add a level $\beta_3$ to $\Z$ with $\map(\beta_3)=(\map(\beta_1)+\map(\beta_2))/2$.
    For $i\in\{1,...,|\Z|\}$ s.t. $u_i=\beta_1$, replace $u_i$
    by $\beta_3$ with probability
    \begin{equation*}
    \frac{\mathbb{P}_s(u_i=\beta_2)}{\mathbb{P}_s(u_i=\beta_1)+\mathbb{P}_s(u_i=\beta_2)},
    \end{equation*}where $\mathbb{P}_s$ is given in~(\ref{eq:def_Boltzmann}); for $i\in\{1,...,|\Z|\}$ s.t. $u_i=\beta_2$, replace $u_i$ by $\beta_3$ with probability
    \begin{equation*}
    \frac{\mathbb{P}_s(u_i=\beta_1)}{\mathbb{P}_s(u_i=\beta_1)+\mathbb{P}_s(u_i=\beta_2)}.
    \end{equation*}Note that $|\{u_i: u_i=\beta_3, i=1,...,N\}|$ is typically non-zero, i.e., $\beta_3$ tends not to be empty.
\end{itemize}
We call the process of running one of these procedures followed by running L-MCMC a {\em round}.

\textbf{Size- and level-adaptive MCMC: }
SLA-MCMC is conceptually illustrated in the flowchart in Fig.~\ref{fig:SLA-MCMC}. It has four stages, and in each stage we will run L-MCMC for several super-iterations; we denote the execution of L-MCMC for $r$ super-iterations by L($r$). The parameters $r_1,r_2,r_3,r_{4a}$, and $r_{4b}$ are the number of super-iterations used in Stages~1 through~4, respectively. The choice of these parameters reflects a trade-off between runtime and estimation quality.

In Stage~1, SLA-MCMC uses a fixed-size adaptive reproduction alphabet $\Z$ to tentatively estimate the signal.
The initial point of Stage~1 is obtained in the same way as L-MCMC.
After Stage~1, the initial point and temperature offset for each instance of L-MCMC correspond to the respective outputs of the previous instance of L-MCMC.
If the source is discrete-valued and $|\Z|>|\X|$ in Stage~1, then multiple levels in the output $\Z$ of Stage~1 may correspond to a single level in $\X$.
To alleviate this problem,  in Stage~2 we merge levels closer than $T=\mathcal{I}_{R_{\map}}/\left(K_1\times(|\Z|-1)\right)$, where $K_1$ is a parameter.

However, $|\Z|$ might still be larger than needed; hence in Stage~3 we tentatively merge the closest adjacent levels. The criterion $D1$ evaluates whether the current objective function is lower (better) than in the previous round;
we do not leave Stage~3 until $D1$ is violated. Note that if $|\X|>|\Z|$ (this always holds for continuous-valued signals), then ideally SLA-MCMC should not merge any levels in Stage~3, because the objective function would increase if we merge any levels.

Define the outlier set $S=\{x_i: x_i\notin R_{\map}, i=1,...,N\}$. Under Condition~\ref{cond:tech1}, $S$ might be small or even empty.
When $S$ is small, L-MCMC might not assign levels to represent the entries of $S$.
To make SLA-MCMC more robust to outliers, in Stage~4a we add empty levels outside the range $R_{\map}$ and then allow L-MCMC to change entries of $u$ to the new levels during Gibbs sampling; we call this
{\em populating} the new levels.
If a newly added outside level is not populated, then we remove it from $\Z$. Seeing that the optimal mapping $\map_{opt}$ in L-MCMC tends not to map symbols to levels with low population, we
consider a criterion $D2$ where we will
will add an outside upper (lower) level if the population of the current upper (lower) level is smaller than $N/(K_2|\Z|)$, where $K_2$ is a parameter.
That is, the criterion $D2$ is violated if both populations of the current upper and lower levels are sufficient (at least $N/(K_2|\Z|)$); in this case we do not need to add outside levels because $\map_{opt}$ will map some of the current levels to represent the entries in $S$.
The criterion $D3$ is violated if all levels added outside are not populated by the end of the round. SLA-MCMC keeps adding levels outside $R_{\map}$ until it is wide enough to cover most of the entries of $x$.

Next, SLA-MCMC considers adding levels inside $R_{\map}$ (Stage~4b). If the signal is discrete-valued, this stage should stop when $|\Z|=|\X|$.
Else, for continuous-valued signals SLA-MCMC can add levels until the runtime expires.

In practice, SLA-MCMC runs L-MCMC at most a constant number of times, and the computational complexity is in the same order of L-MCMC, i.e., $O(rMN|\Z|)$. On the other hand, SLA-MCMC allows varying
$|\Z|$, which often improves the estimation quality.

\subsection{Mixing}\label{sec:mix}

Donoho proved for the scalar channel setting that $x_{KS}$ is sampled from the posterior $\mathbb{P}_{X|Y}(x|y)$~\cite{DonohoKolmogorov}.
Seeing that the Gibbs sampler used by MCMC (cf.\ Section~\ref{sec:B-MCMC}) generates random samples, and the
outputs of our algorithm will be different if its random number generator is initialized with different {\em random seeds}, we speculate that running SLA-MCMC several times
will also yield independent samples from the posterior, where we note that
the runtime grows linearly in the number of times that we run SLA-MCMC.
By mixing (averaging over) several outputs of SLA-MCMC, we obtain $\widehat{x}_{\mbox{avg}}$, which may
have lower squared error w.r.t.\ the true $x$ than the average squared error obtained by a single SLA-MCMC output.
Numerical results suggest that mixing indeed reduces the MSE (cf.\ Fig.~\ref{fig:MUnif_algos});
this observation suggests that mixing the outputs of multiple algorithms, including running a
random reconstruction algorithm several times, may reduce the squared error.

\section{Numerical results} \label{sec:numerical}

In this section, we demonstrate that SLA-MCMC is comparable and in many cases better than existing algorithms in reconstruction quality,
and that SLA-MCMC is applicable when $M>N$. Additionally, some numerical evidence is provided to justify Conjecture~\ref{conj:double_MMSE} in Section~\ref{sec:conjecture}.
Then, the advantage of SLA-MCMC in estimating low-complexity signals is demonstrated. Finally, we compare B-MCMC, L-MCMC, and SLA-MCMC performance.

We implemented SLA-MCMC in Matlab\footnote{A toolbox that runs the simulations in this paper is available at http://people.engr.ncsu.edu/dzbaron/software/UCS\_BaronDuarte/} and tested it using several stationary ergodic sources.
Except when noted, for each source, signals $x$ of length $N=10000$ were generated.
Each such $x$ was multiplied by a Gaussian random matrix $\Phi$ with normalized columns and corrupted by i.i.d. Gaussian measurement noise $z$.
Except when noted, the number of measurements $M$ varied between 2000 and 7000.
The noise variance $\sigma_Z^2$ was selected to ensure that the signal-to-noise ratio (SNR) was $5$ or $10$~dB; SNR was defined as
$\displaystyle \mbox{SNR}=10\log_{10}\left[(N E[x^2])/(M\sigma_Z^2)\right]$.
According to Section~\ref{sec:compressor}, the context depth $q=o(\log(N))$, where the base of the logarithm is the alphabet size; using typical values such as $N=10000$ and $|\Z|=10$, we have $\log(N)=4$ and set $q=2$. While larger $q$ will slow down the algorithm, it might be necessary to increase $q$ when $N$ is larger. The number of super-iterations in different stages of SLA-MCMC $r_{1}=50$ and $r_2=r_3=r_{4a}=r_{4b}=10$, the maximum total number of super-iterations to be $240$, the initial number of levels $|\Z|=7$, and the tuning parameter from Section~\ref{sec:adaptive_size} $K_1,K_2=10$; these parameters seem to work well on an extensive set of numerical experiments.
SLA-MCMC was not given the true alphabet $\X$ for any of the sources presented in this paper; our expectation is that it should adaptively adjust $|\Z|$ to match $|\X|$.
The final estimate $\widehat{x}_{\mbox{avg}}$ of each signal was obtained by averaging over the outputs $\widehat{x}$ of $5$
runs of SLA-MCMC, where in each run we initialized the random number generator with another random seed, cf.\ Section~\ref{sec:mix}. These choices of parameters seemed to provide a reasonable compromise between runtime and estimation quality.

We chose our performance metric as the mean signal-to-distortion ratio (MSDR) defined as
$\displaystyle \mbox{MSDR}=10\log_{10}\left(E[x^2]/\mbox{MSE}\right)$. For each $M$ and SNR, the MSE was obtained after averaging over the squared errors of $\widehat{x}_{\mbox{avg}}$ for 50 draws of $x$, $\Phi$, and $z$.
We compared the performance of SLA-MCMC to
that of ({\em i}) compressive sensing matching pursuit (CoSaMP)~\cite{Cosamp08}, a greedy method;
({\em ii}) gradient projection for sparse reconstruction (GPSR)~\cite{GPSR2007}, an optimization-based method; 
({\em iii}) message passing approaches (for each source, we chose best-matched algorithms between EM-GM-AMP-MOS (EGAM for short)~\cite{EMGMTSP} and turboGAMP (tG for short)~\cite{turboGAMP}); and
({\em iv}) Bayesian compressive sensing~\cite{BCS2008} (BCS).
Note that EGAM~\cite{EMGMTSP} places a Gaussian mixture prior on the signal, and tG~\cite{turboGAMP} builds a prior set including the priors for the signal, the support set of the signal, the channel, and the amplitude structure. Both algorithms learn the parameters of their assumed priors online from the measurements. We compare the computational complexities of the algorithms above in Table~\ref{table:complexity}, where
$L$ bounds the cost of a matrix–-vector multiply with $\Phi$ or the Hermitian transpose of $\Phi$, and $\epsilon$ is a given precision parameter~\cite{Cosamp08}; $r_P,r_E, r_G, r_M$ are the number of GPSR~\cite{GPSR2007}, Expectation Maximization (EM), GAMP~\cite{RanganGAMP2011ISIT}, and model selection~\cite{EMGMTSP} iterations, respectively; $T_1$ and $T_2$ are the average complexities for the EM algorithm and the turbo updating scheme~\cite{turboGAMP}.
\begin{table}[t!]
\caption{\small{Computational complexity}} 
\centering 
\begin{tabular}{c c} 
\hline\hline 
Algorithms & Complexity\\ [0.5ex] 
\hline 
SLA-MCMC & $O(rMN|\Z|)$    \\ 
CoSaMP & $O(L\log\frac{||x||}{\epsilon})$\\
GPSR & $O(r_P MN)$ \\
EGAM & $O(r_M r_E T_1+r_M r_E r_G MN)$ \\
tG & $O(r_E T_2+r_E r_G MN)$ \\ [1ex] 
\hline 
\end{tabular}
\label{table:complexity} 
\end{table}
Because all these algorithms are iterative algorithms and require different number of iterations to converge or reach a satisfactory reconstruction quality, we also report their typical runtimes here. Typical runtimes are $1$ hour (for continuous-valued signals) and $15$ minutes (discrete-valued)
per random seed for SLA-MCMC, $30$ minutes for EGAM~\cite{EMGMTSP} and tG~\cite{turboGAMP}, and $10$ minutes for CoSaMP~\cite{Cosamp08} and GPSR~\cite{GPSR2007} on an Intel(R) Core(TM) i7 CPU 860 @ 2.8GHz with 16.0GB RAM running 64 bit Windows 7. The performance of BCS was roughly 5~dB below SLA-MCMC results. Hence, BCS results are not shown in the sequel. We emphasize that algorithms that use training data (such as dictionary learning)~\cite{Ramirez2011,AharoEB_KSVD,Mairal2008,Zhoul2011} will find our problem size $N=10000$ too large, because they need a training set that has more than $N$ signals. On the other hand, SLA-MCMC does not need to train itself on any training set, and hence is advantageous.

Among these baseline algorithms designed for i.i.d. signals, GPSR~\cite{GPSR2007} and EGAM~\cite{EMGMTSP} only need $y$ and $\Phi$, and CoSaMP~\cite{Cosamp08} also needs the number of non-zeros in $x$. Only tG~\cite{turboGAMP} is designed for non-i.i.d. signals; however, it must be aware of the probabilistic model of the source. Finally, GPSR~\cite{GPSR2007} performance was similar to that of CoSaMP~\cite{Cosamp08} for all sources considered in this section, and thus is not plotted.

\subsection{Performance on discrete-valued sources}
\begin{figure}[t]
\begin{center}
\includegraphics[width=80mm]{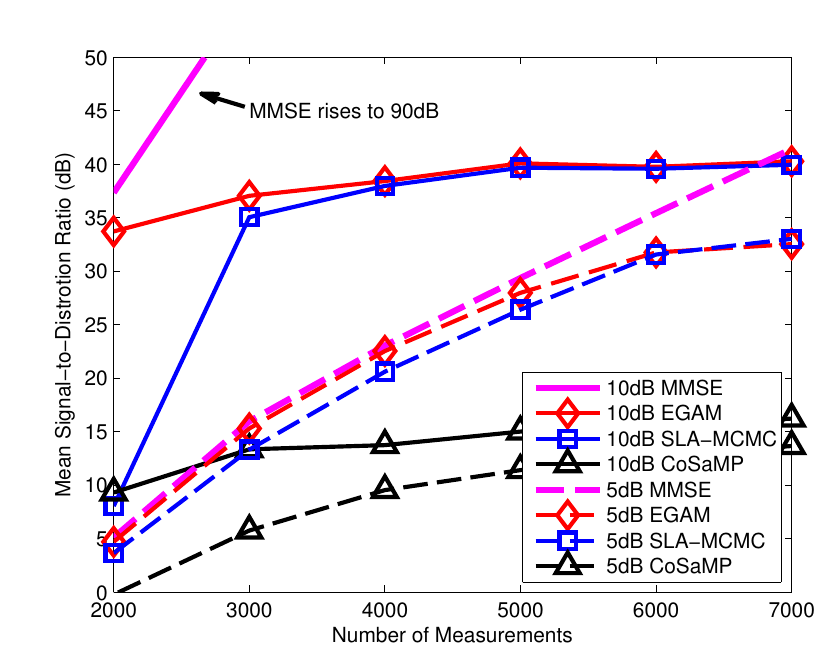}
\end{center}
\vspace*{-5mm}
\caption{{\small\sl SLA-MCMC, EGAM, and CoSaMP estimation results for a source with i.i.d. Bernoulli entries with non-zero probability of $3\%$ as
a function of the number of Gaussian random measurements $M$ for different SNR values ($N=10000$).}
\label{fig:Ber}}
\vspace*{-5mm}
\end{figure}

{\bf Bernoulli source:} We first present results for an i.i.d. Bernoulli source. The Bernoulli source followed the distribution $f_X(x)=0.03\delta(x-1)+0.97\delta(x)$, where $\delta(\cdot)$ is the Dirac delta function. Note that SLA-MCMC did not know $\X=\{0,1\}$ and had to estimate it on the fly.
We chose EGAM~\cite{EMGMTSP} for message passing algorithms because it fits the signal with Gaussian mixtures (GM),
which can accurately characterize signals from an i.i.d. Bernoulli source.
The resulting MSDRs for SLA-MCMC, EGAM~\cite{EMGMTSP}, and CoSaMP~\cite{Cosamp08} are plotted in Fig.~\ref{fig:Ber}.
We can see that when $\mbox{SNR}=5~\mbox{dB}$, EGAM~\cite{EMGMTSP} approaches the MMSE~\cite{ZhuBaronCISS2013} performance for low to medium $M$; although SLA-MCMC is often worse than EGAM~\cite{EMGMTSP}, it is within $3$~dB of the MMSE performance.
This observation that SLA-MCMC approaches the MMSE for $\mbox{SNR}=5~\mbox{dB}$ partially substantiates Conjecture~\ref{conj:double_MMSE} in Section~\ref{sec:conjecture}. When $\mbox{SNR}=10~\mbox{dB}$, SLA-MCMC is comparable to EGAM~\cite{EMGMTSP} when $M\geq 3000$. CoSaMP~\cite{Cosamp08} has worse MSDR.

\begin{figure}[t]
\begin{center}
\includegraphics[width=80mm]{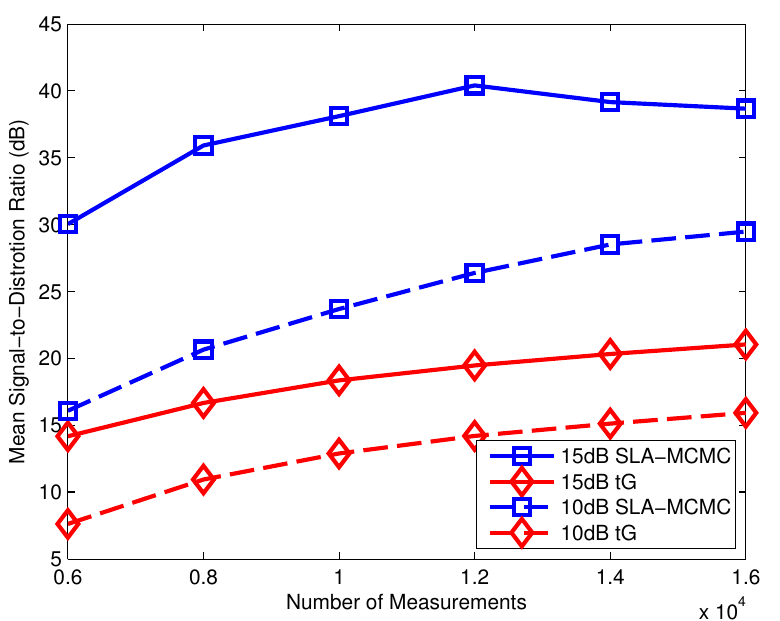}
\end{center}
\vspace*{-5mm}
\caption{{\small\sl SLA-MCMC and tG estimation results for a dense two-state Markov source with non-zero entries drawn from a Rademacher ($\pm 1$) distribution as
a function of the number of Gaussian random measurements $M$ for different SNR values ($N=10000$).}
\label{fig:denseMRad}}
\vspace*{-5mm}
\end{figure}

{\bf Dense Markov-Rademacher source:}
Considering that most algorithms are designed for i.i.d. sources, we now illustrate the performance of SLA-MCMC on non-i.i.d. sources by simulating a dense Markov-Rademacher (MRad for short) source. The non-zero entries of the dense MRad signal were generated by a two-state Markov state machine (non-zero and zero states). The transition from zero to non-zero state for adjacent entries had probability $\mathbb{P}_{01}=\frac{3}{70}$, while the transition from non-zero to zero state for adjacent entries had probability $\mathbb{P}_{10}=0.10$; these parameters yielded $30\%$ non-zero entries on average. The non-zeros were drawn from a Rademacher distribution, which took values $\pm1$ with equal probability.
With such denser signals, we may need to take more measurements and/or require higher SNRs to achieve similar performance to previous examples.
The number of measurements varied from $6000$ to $16000$, with $\mbox{SNR}=10~\mbox{and}~15~\mbox{dB}$.
Although tG~\cite{turboGAMP} does not provide an option that accurately characterize the MRad source, we still chose
to compare against its performance because it is applicable to non-i.i.d. signals.
The MSDRs for SLA-MCMC and tG~\cite{turboGAMP} are plotted in Fig.~\ref{fig:denseMRad}. CoSaMP~\cite{Cosamp08} performs poorly as it is designed for sparse signal recovery, and its results are not shown. Although tG~\cite{turboGAMP} is designed for non-i.i.d. sources, it is nonetheless outperformed by SLA-MCMC.
This example shows that SLA-MCMC reconstructs non-i.i.d. signals well and is applicable to general linear inverse problems. However, recall that the computational complexity of SLA-MCMC is $O(rMN|\Z|)$. Hence, despite the appealing performance of SLA-MCMC shown in this example, we will suffer from high computational time when we have to apply SLA-MCMC in the case when $M>N$.

\subsection{Performance on continuous sources}\label{sec:bvsub}

We now discuss the performance of SLA-MCMC in estimating continuous sources.

\begin{figure}[t]
\begin{center}
\includegraphics[width=80mm]{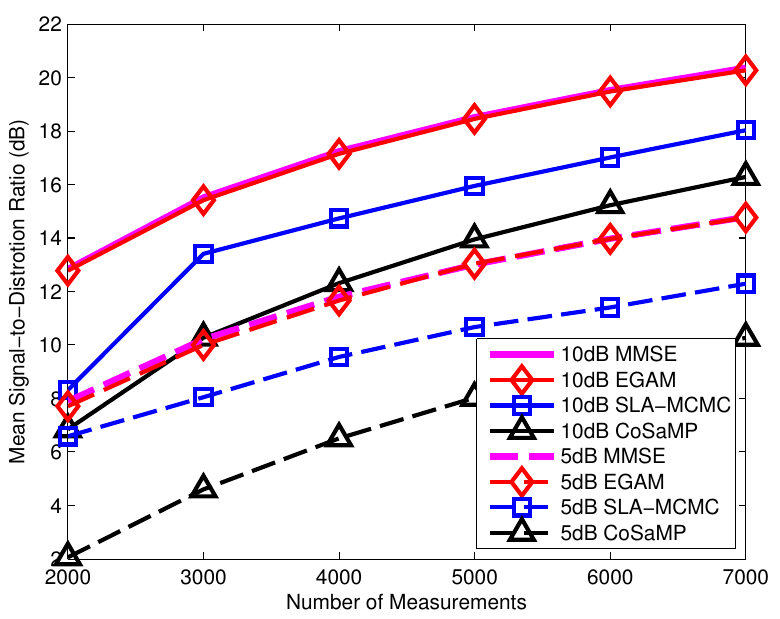}
\end{center}
\vspace*{-5mm}
\caption{{\small\sl SLA-MCMC, EGAM, and CoSaMP estimation results for an i.i.d. sparse Laplace source as
a function of the number of Gaussian random measurements $M$ for different SNR values ($N=10000$).}
\label{fig:sparseL}}
\vspace*{-5mm}
\end{figure}

{\bf Sparse Laplace (i.i.d.) source:} For unbounded continuous-valued signals, which do not adhere to Condition~\ref{cond:tech1}, we simulated an i.i.d. sparse Laplace source following the distribution $f_X(x)=0.03\mathcal{L}(0,1)+0.97\delta(x)$, where $\mathcal{L}(0,1)$ denotes a Laplacian distribution with mean zero and variance one.
We chose EGAM~\cite{EMGMTSP} for message passing algorithms because it fits the signal with GM,
which can accurately characterize signals from an i.i.d. sparse Laplace source.
The MSDRs for SLA-MCMC, EGAM~\cite{EMGMTSP}, and CoSaMP~\cite{Cosamp08} are plotted in Fig.~\ref{fig:sparseL}.
We can see that EGAM~\cite{EMGMTSP} approaches the MMSE~\cite{ZhuBaronCISS2013} performance in all settings;
SLA-MCMC outperforms CoSaMP~\cite{Cosamp08}, while it is approximately $2$~dB worse than the MMSE.
Recall from Conjecture~\ref{conj:double_MMSE} that we expect to achieve twice the MMSE, which is approximately $3$~dB below the signal-to-distortion ratio of MMSE, and thus SLA-MCMC performance is reasonable.
This example of SLA-MCMC performance approaching the MMSE further substantiates Conjecture~\ref{conj:double_MMSE}.

\begin{figure}[t]
\begin{center}
\includegraphics[width=80mm]{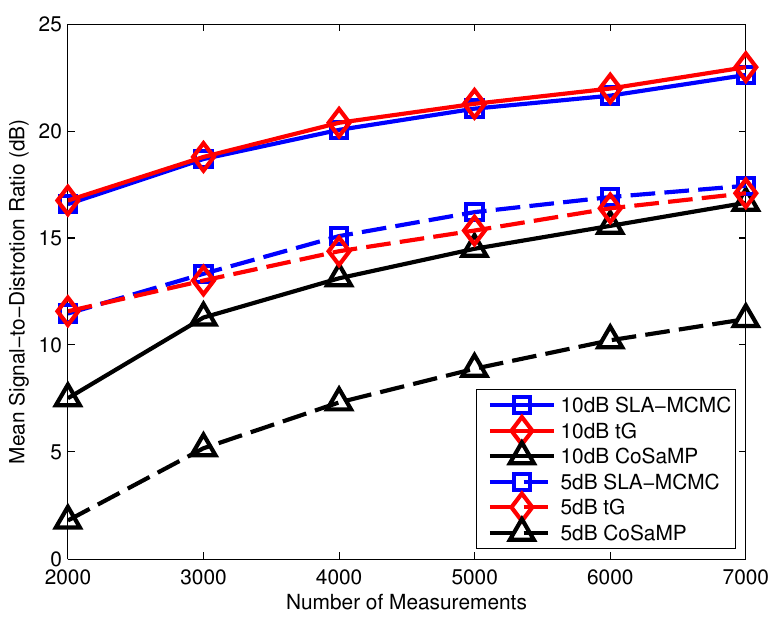}
\end{center}
\vspace*{-5mm}
\caption{{\small\sl SLA-MCMC, tG, and CoSaMP estimation results for a two-state Markov source with non-zero entries drawn from
a uniform distribution $U[0,1]$ as a function of the number of Gaussian random measurements $M$ for different SNR values ($N=10000$).}
\label{fig:MUnif}}
\vspace*{-5mm}
\end{figure}

{\bf Markov-Uniform source:} For bounded continuous-valued signals, which adhere to Condition~\ref{cond:tech1}, we simulated a Markov-Uniform (MUnif for short) source, whose non-zero entries were generated
by a two-state Markov state machine (non-zero and zero states) with $\mathbb{P}_{01}=\frac{3}{970}$ and $\mathbb{P}_{10}=0.10$; these parameters yielded $3\%$ non-zeros entries on average. The non-zero entries were drawn from
a uniform distribution between $0$ and $1$.
We chose tG with Markov support and GM model options~\cite{turboGAMP} for message passing algorithms.
We plot the resulting MSDRs for SLA-MCMC, tG~\cite{turboGAMP}, and CoSaMP~\cite{Cosamp08} in Fig.~\ref{fig:MUnif}.
We can see that the CoSaMP~\cite{Cosamp08} lags behind in MSDR. The SLA-MCMC curve is close to that of tG~\cite{turboGAMP} when $\mbox{SNR}=10~\mbox{dB}$, and it is slightly better than tG~\cite{turboGAMP} when $\mbox{SNR}=5~\mbox{dB}$.

When the signal model is known, the message passing approaches EGAM~\cite{EMGMTSP} and tG~\cite{turboGAMP} achieve quite low MSE's, because they can get close to the Bayesian MMSE. Sometimes the model is only known imprecisely, and SLA-MCMC can improve over message passing; for example, it is better than tG~\cite{turboGAMP} in estimating MUnif signals (Fig.~\ref{fig:MUnif}), because tG~\cite{turboGAMP} approximates the uniformly distributed non-zeros by GM.

\subsection{Comparison between discrete and continuous sources}

When the source is continuous (Figs.~\ref{fig:sparseL} and~\ref{fig:MUnif}), SLA-MCMC might be worse than the existing message passing approaches (EGAM~\cite{EMGMTSP} and tG~\cite{turboGAMP}).
One reason for the under-performance of SLA-MCMC is the $3$~dB gap of Conjecture~\ref{conj:double_MMSE}. The second reason is that SLA-MCMC can only assign finitely many levels to approximate continuous-valued signals, leading to under-representation of the signal. However, when it comes to discrete-valued signals that have finite size alphabets (Figs.~\ref{fig:Ber} and~\ref{fig:denseMRad}), SLA-MCMC is comparable to and in many cases better than existing algorithms. Nonetheless, we observe in the figures that SLA-MCMC is far from the state-of-the-art when the SNR is high and measurement rate is low. Additionally, the dense MRad source in Fig.~\ref{fig:denseMRad} has only a limited number of discrete levels and may not provide a general enough example.

\begin{figure}[t]
\begin{center}
\includegraphics[width=80mm]{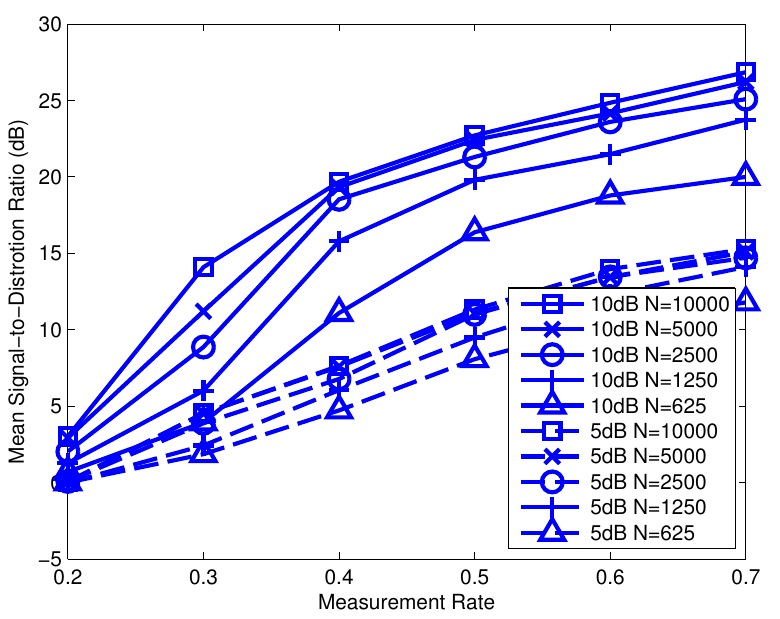}
\end{center}
\vspace*{-5mm}
\caption{{\small\sl SLA-MCMC estimation results for a four-state Markov switching source as
a function of the measurement rate $R$ for different SNR values and signal lengths. Existing CS algorithms fail at reconstructing this signal, because this source is not sparse.}
\label{fig:M4}}
\vspace*{-5mm}
\end{figure}

\subsection{Performance on low-complexity signals}
SLA-MCMC promotes low complexity due to the complexity-penalized term in the objective function~(\ref{eq:MCMC_energy}). Hence, it tends to perform well for signals with low complexity such as the signals in
Figs.~\ref{fig:Ber} and~\ref{fig:denseMRad} (note that the Bernoulli signal is sparse while the MRad signal is denser).
In this subsection, we simulated a non-sparse low-complexity signal. We show that complexity-penalized approaches such as SLA-MCMC might estimate low-complexity signals well.

{\bf Four-state Markov source:} To evaluate the performance of SLA-MCMC for discrete-valued non-i.i.d.\ and non-sparse signals, we examined a four-state Markov source (Markov4 for short) that generated the pattern $+1,+1,-1,-1,+1,+1,-1,-1\ldots$
with 3\% errors in state transitions, resulting in the signal switching from $-1$ to $+1$ or
vice versa either too early or too late. Note that the reconstruction algorithm did not know that this source is a binary source. While it is well known that sparsity-promoting
recovery algorithms~\cite{turboGAMP,Cosamp08,GPSR2007}
can recover sparse sources from linear measurements,
the aforementioned switching source is not sparse in conventional sparsifying bases (e.g., Fourier, wavelet, and discrete cosine transforms), rendering such sparsifying transforms not applicable. Signals generated by this Markov source can be sparsified using an averaging analysis matrix~\cite{CandesCSdictonary2011} whose diagonal and first three lower sub-diagonals are filled with $+1$, and all other entries are $0$; this transform yields $6\%$ non-zeros in the sparse coefficient vector.
However, even if this matrix had been known {\em a priori}, existing algorithms based on analysis sparsity~\cite{CandesCSdictonary2011} did not perform satisfactorily, yielding mean signal-to-distortion ratios below $5$~dB.
Thus, we did not include the results for these baseline algorithms in Fig.~\ref{fig:M4}.
On the other hand, Markov4 signals have low complexity in the time domain, and hence, SLA-MCMC successfully reconstructed Markov4 signals with reasonable quality even when $M$ was relatively small.
This Markov4 source highlights the special advantage of our approach in reconstructing low-complexity signals.

The MSDRs for shorter Markov4 signals are also plotted in Fig.~\ref{fig:M4}. We can see that SLA-MCMC performs better when the signal to be reconstructed is longer. Indeed, SLA-MCMC needs a signal that is long enough to learn the statistics of the signal.

\subsection{Performance on real world signals}
\begin{figure}[t]
\begin{center}
\includegraphics[width=80mm]{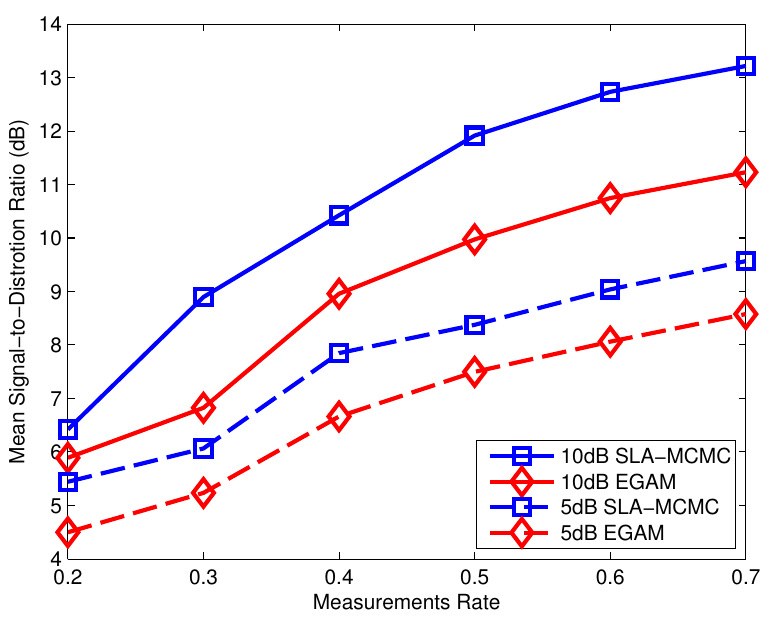}
\end{center}
\vspace*{-5mm}
\caption{{\small\sl SLA-MCMC and EGAM estimation results for a Chirp signal as a function of the measurement rate $R$ for different SNR values ($N=9600$).}
\label{fig:Chirp}}
\vspace*{-5mm}
\end{figure}

Our experiments up to this point use synthetic signals, where SLA-MCMC has shown comparable and in many cases better results than existing algorithms. This subsection evaluates how well SLA-MCMC reconstructs a real world signal. We use the ``Chirp'' sound clip from Matlab: we cut a consecutive part with length 9600 out of the ``Chirp'' (denoted by $x$) and performed a short-time discrete cosine transform (DCT) with window size, number of DCT points, and hop size all being 32.
Then we vectorized the resulting short-time DCT coefficients matrix to form a coefficient vector $\theta$ of length 9600. By denoting the short-time DCT matrix by $W^{-1}$, we have $\theta=W^{-1}x$. Therefore, we can rewrite~\eqref{eq:def_y} as $y=A\theta +z$,
where $A=\Phi W$. We want to reconstruct $\theta$ from the measurements $y$ and the matrix $A$. After we obtain the estimate $\widehat{\theta}$, we obtain the estimated signal by $\widehat{x}=W\widehat{\theta}$.
Although the coefficient vector $\theta$ may exhibit some type of memory, it is not readily modeled in closed form, and so we cannot provide a valid model for tG~\cite{turboGAMP}. Instead, we use EGAM~\cite{EMGMTSP} as our benchmark algorithm. We do not compare to CoSaMP~\cite{Cosamp08} because it falls behind in performance as we have seen from other examples. The MSDRs for SLA-MCMC and EGAM~\cite{EMGMTSP} are plotted in Fig.~\ref{fig:Chirp}, where SLA-MCMC outperforms EGAM by 1--2 dB.

\begin{figure}[t]
\begin{center}
\includegraphics[width=80mm]{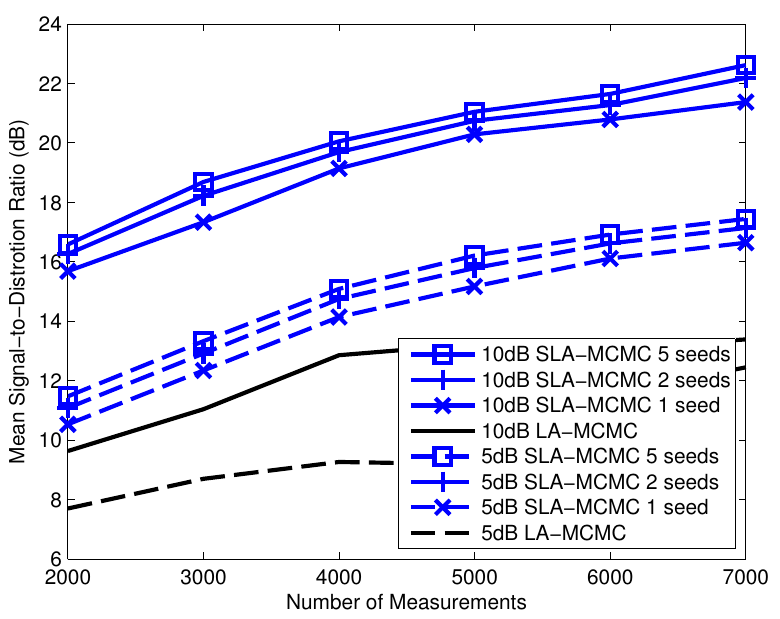}
\end{center}
\vspace*{-5mm}
\caption{{\small\sl SLA-MCMC with different number of random seeds and L-MCMC  estimation results for the Markov-Uniform source described in Fig.~\ref{fig:MUnif} as a function of the number of Gaussian random measurements $M$ for different SNR values ($N=10000$).}
\label{fig:MUnif_algos}}
\vspace*{-5mm}
\end{figure}

\subsection{Comparison of B-MCMC, L-MCMC, and SLA-MCMC}
We compare the performance of B-MCMC, L-MCMC, and SLA-MCMC with different numbers of seeds (cf.\ Section~\ref{sec:mix}) by examining the MUnif source (cf.\ Section~\ref{sec:bvsub}). We ran B-MCMC with the fixed uniform alphabet $\replevels_F$ in~(\ref{eq:def:replevels2}) with $|\replevels_F|=10$ levels. L-MCMC was initialized in the same way as Stage~1 of SLA-MCMC. B-MCMC and L-MCMC ran for $100$ super-iterations before outputting the estimates; this number of super-iterations was sufficient because it was greater than $r_1=50$ in Stage~1 of SLA-MCMC. The results are plotted in Fig.~\ref{fig:MUnif_algos}.
B-MCMC did not perform well given the $\replevels_F$ in~(\ref{eq:def:replevels2}) and is not plotted.
We can see that SLA-MCMC outperforms L-MCMC. Averaging over more seeds provides an increase of $1$~dB in MSDR.\footnote{For other sources, we observed an increase in MSDR of up to $2$~dB.} It is likely that averaging over more seeds with each seed running fewer super-iterations will decrease the squared error. We leave the optimization of the number of seeds and the number of super-iterations in each seed for future work.
Finally, we tried a ``good" reproduction alphabet in B-MCMC, $\displaystyle\widetilde{\replevels}_F=\frac{1}{|\replevels_F|-1/2}\{0,...,|\replevels_F|-1\}$, and the results were close to those of SLA-MCMC. Indeed, B-MCMC is quite sensitive to the reproduction alphabet, and Stages~$2$--$4$ of SLA-MCMC find a good set of levels.
Example output levels $\map(\Z)$ of SLA-MCMC were: $\{-0.001,0.993\}$ for Bernoulli signals, $\{-0.998,0.004,1.004\}$ for dense MRad signals, $21$ levels spread in the range $[-3.283,4.733]$ for i.i.d. sparse Laplace signals, $22$ levels spread in the range $[-0.000,0.955]$ for MUnif signals, and $\{-1.010,0.996\}$ for Markov4 signals; we can see that SLA-MCMC adaptively adjusted $|\Z|$ to match $|\X|$ so that these levels represented each signal well. Also, we can see from Figs.~\ref{fig:Ber}--\ref{fig:sparseL} that SLA-MCMC did not perform well in the low measurements and high SNR setting, which was due to mismatch between $|\Z|$ and $|\X|$.

\section{Conclusions} \label{sec:conclusions}

This paper provides universal algorithms for signal estimation from linear measurements. Here, universality
denotes the property that the algorithm need not be informed of the probability
distribution for the recorded signal prior to acquisition; rather, the algorithm
simultaneously builds estimates both of the observed signal and its distribution.
Inspired by the Kolmogorov sampler~\cite{DonohoKolmogorov} and motivated by the
need for a computationally tractable framework, our contribution
focuses on stationary ergodic signal sources and relies on a MAP estimation algorithm. The algorithm is then implemented via a MCMC formulation that is proven to be convergent in the limit of infinite computation.
We reduce the computation and improve the estimation quality of the proposed algorithm by adapting the reproduction alphabet to match
the complexity of the input signal. Our experiments have shown that the performance of the proposed algorithm
is comparable to and in many cases better than existing
algorithms,
particularly for low-complexity sources that do not exhibit standard sparsity or compressibility.

As we were finishing this paper, Jalali and Poor~\cite{JalaliPoor2014} have independently shown that our formulation~(\ref{eq:MCMC_energy}) also provides an implementable version of R{\'e}nyi entropy minimization. Their theoretical findings further motivate our proposed universal MCMC formulation.
\renewcommand\thesection{Appendix \Alph{section}}
\setcounter{section}{0}

\section{Proof of Theorem~\ref{th:conv}}
\label{ap:th:conv}

Our proof mimics a very similar proof presented in~\cite{Jalali2008,Jalali2012} for lossy source coding; we include all details for completeness. The proof technique relies on mathematical properties of non-homogeneous (e.g., time-varying) Markov Chains (MCs)~\cite{Bremaud1999}. Through the proof, $\ss \triangleq (\replevels_F)^N$ denotes the state space of the MC of codewords generated by Algorithm~\ref{alg:MCMC}, with size $|\ss| = |\replevels_F|^N$. We define a stochastic transition matrix $P_{(t)}$ from $\ss$ to itself given by the Boltzmann distribution for super-iteration $t$ in Algorithm~\ref{alg:MCMC}. Similarly, $\pi_{(t)}$ defines the stable-state distribution on $\ss$ for $P_{(t)}$, satisfying $\pi_{(t)} P_{(t)} = \pi_{(t)}$.

\begin{DEFI}~\cite{Bremaud1999}
{\em Dobrushin's ergodic coefficient} of a MC transition matrix $P$ is denoted by $\xi(P)$ and defined as
$\displaystyle \xi(P) \triangleq \max_{1 \le i,j \le N} \frac{1}{2}\|p_i-p_j\|_1$,
where $p_{i}$ denotes row $i$ of $P$.
\end{DEFI}
From the definition, $0 \le \xi(P) \le 1$. Moreover, the ergodic coefficient can be rewritten as
\begin{align}
\xi(P) = 1-\min_{1 \le i,j \le N} \sum_{k=1}^N \min(p_{ik},p_{jk}),
\label{eq:ergcoef}
\end{align}
where $p_{ij}$ denotes the entry of $P$ at row $i$ and column $j$.

We group the product of transition matrices
across super-iterations as $P_{(t_1\to t_2)} = \prod_{t=t_1}^{t_2}P_{(t)}$.
There are two common characterizations for the stable-state behavior of a
non-homogeneous MC.
\begin{DEFI}~\cite{Bremaud1999}
A non-homogeneous MC is called {\em weakly ergodic} if for any distributions
$\eta$ and $\nu$ over the state space $\ss$, and any $t_1 \in \mathbb{N}$,
$\displaystyle {\lim\sup}_{t_2 \to \infty} \|\eta P_{(t_1\to t_2)}-\nu P_{(t_1\to t_2)}\|_1 = 0$,
where $\|\cdot\|_1$ denotes the $\ell_1$ norm.
Similarly, a non-homogeneous MC is called {\em strongly ergodic} if there
exists a distribution $\pi$ over the state space $\ss$ such that for any
distribution $\eta$ over $\ss$, and any $t_1 \in \mathbb{N}$,
$\displaystyle {\lim\sup}_{t_2 \to \infty} \|\eta P_{(t_1 \to t_2)}-\pi\|_1 = 0$.
We will use the following two theorems from~\cite{Bremaud1999} in our proof.
\end{DEFI}
\begin{THEO}~\cite{Bremaud1999}
A MC is weakly ergodic if and only if there exists a sequence of integers
$0 \le t_1 \le t_2 \le \ldots$ such that $\displaystyle\sum_{i=1}^\infty \left(1-\xi\left(P_{(t_i \to t_{i+1})}\right)\right) = \infty$.
\label{th:block}
\end{THEO}
\begin{THEO}~\cite{Bremaud1999}
Let a MC be weakly ergodic. Assume that there exists a sequence of
probability distributions $\{\pi_{(t)}\}_{i=1}^\infty$ on the state space $\ss$
such that $\pi_{(t)} P_{(t)} = \pi_{(t)}$. Then the MC is strongly ergodic if $\displaystyle\sum_{t=1}^\infty \|\pi_{(t)}-\pi_{(t+1)}\|_1 < \infty$.
\label{th:weak}
\end{THEO}

The rest of proof is structured as follows. First, we show that the sequence
of stable-state distributions for the MC used by Algorithm~\ref{alg:MCMC}
converges to a uniform distribution over the set of sequences that minimize
the energy function as the iteration count $t$ increases. Then, we show using
Theorems~\ref{th:block} and~\ref{th:weak} that the non-homogeneous MC
used in Algorithm~\ref{alg:MCMC} is strongly ergodic, which by the definition
of strong ergodicity implies that Algorithm~\ref{alg:MCMC} always converges
to the stable distribution found above. This implies that the outcome of
Algorithm~\ref{alg:MCMC} converges to a minimum-energy solution as
$t \to \infty$, completing the proof of Theorem~\ref{th:conv}.

We therefore begin by finding the stable-state distribution for the non-homogeneous
MC used by Algorithm~\ref{alg:MCMC}. At each super-iteration $t$, the distribution
defined as
\begin{equation}
\begin{split}
\pi_{(t)}(w) &\triangleq \frac{\exp(-s_t\Psi^{H_q}(w))}{\sum_{z \in \ss} \exp(-s_t\Psi^{H_q}(z))}\\
&= \frac{1}{\sum_{z \in \ss} \exp(-s_t(\Psi^{H_q}(z)-\Psi^{H_q}(w)))} \label{eq:pidist}
\end{split}
\end{equation}
satisfies $\pi_{(t)}P_{(t)} = \pi_{(t)}$, cf.~(\ref{eqn:Gibbs}). We can show that the
distribution $\pi_{(t)}$ converges to a uniform distribution over the set of sequences
that minimize the energy function, i.e.,
\begin{align}
\lim_{t \to \infty} \pi_{(t)}(w) = \left\{\begin{array}{cl}0 & w \notin \mathcal{H}, \\ \frac{1}{|\mathcal{H}|} & w \in \mathcal{H},\end{array}\right.
\label{eq:stable}
\end{align}
where $\mathcal{H} = \{w \in \ss~\textrm{s.t.}~\Psi^{H_q}(w) = \min_{z \in \ss} \Psi^{H_q}(z)\}$.
To show (\ref{eq:stable}), we will show that $\pi_{(t)}(w)$ is increasing for
$w \in \mathcal{H}$ and eventually decreasing for $w \in \mathcal{H}^C$.
Since for $w \in \mathcal{H}$ and $\widetilde{w} \in \ss$ we have
$\Psi^{H_q}(\widetilde{w})-\Psi^{H_q}(w) \ge 0$,  for $t_1 < t_2$ we have
\begin{equation*}
\begin{split}
&\sum_{\widetilde{w} \in \ss} \exp(-s_{t_1}(\Psi^{H_q}(\widetilde{w})-\Psi^{H_q}(w)))\\
\ge &\sum_{\widetilde{w} \in \ss} \exp(-s_{t_2}(\Psi^{H_q}(\widetilde{w})-\Psi^{H_q}(w))),
\end{split}
\end{equation*}
which together with (\ref{eq:pidist}) implies $\pi_{(t_1)}(w) \le \pi_{(t_2)}(w)$. On the other hand, if $w \in \mathcal{H}^C$, then we obtain (\ref{eq:hlower}).
\begin{figure*}
\vspace*{-5mm}
\begin{align}
\pi_{(t)}(w) = \left[\sum_{\widetilde{w}: \Psi^{H_q}(\widetilde{w}) \ge \Psi^{H_q}(w)} \exp(-s_t(\Psi^{H_q}(\widetilde{w})-\Psi^{H_q}(w)))+\sum_{\widetilde{w}: \Psi^{H_q}(\widetilde{w}) < \Psi^{H_q}(w)} \exp(-s_t(\Psi^{H_q}(\widetilde{w})-\Psi^{H_q}(w)))\right]^{-1} \label{eq:hlower}
\end{align}
\end{figure*}
For sufficiently large $s_t$, the denominator of (\ref{eq:hlower}) is dominated
by the second term, which increases when $s_t$ increases, and therefore
$\pi_{(t)}(w)$ decreases for $w \in \mathcal{H}^C$ as $t$ increases. Finally,
since all sequences $w \in \mathcal{H}$ have the same energy $\Psi^{H_q}(w)$,
it follows that the distribution is uniform over the symbols in $\mathcal{H}$.

Having shown convergence of the non-homogenous MC's stable-state distributions,
we now show that the non-homogeneous MC is strongly ergodic. The transition
matrix $P_{(t)}$ of the MC at iteration $t$ depends on the temperature $s_t$ in
(\ref{eq:st}) used within Algorithm~\ref{alg:MCMC}. We first show that the MC
used in Algorithm~\ref{alg:MCMC} is weakly ergodic via Theorem~\ref{th:block}; the
proof of the following Lemma is given in~\ref{app:lemm:ergcoefbound}.
\begin{LEMM}
The ergodic coefficient of $P_{(t)}$ for any $t\ge 0$ is upper bounded by
$\displaystyle \xi\left(P_{(t)}\right) \le 1-\exp(-s_tN\Delta_q)$,
where $\Delta_q$ is defined in (\ref{eq:Deltaq}).
\label{lemm:ergcoefbound}
\end{LEMM}
We note in passing that Condition~\ref{cond:tech1} ensures that $\Delta_q$ is
finite. Using Lemma~\ref{lemm:ergcoefbound} and (\ref{eq:st}), we can evaluate
the sum given in Theorem~\ref{th:block} as
\begin{align*}
\sum_{j=1}^\infty\left(1-\xi\left(P_{(j)}\right)\right) \ge \sum_{j=1}^\infty\exp(-s_jN\Delta_q)
= \sum_{j=1}^\infty\frac{1}{j^{1/c}} = \infty,
\end{align*}
and so the non-homogeneous MC defined by $\{P_{(t)}\}_{t=1}^\infty$ is
weakly ergodic. Now we use Theorem~\ref{th:weak} to show that the MC is
strongly ergodic by proving that
$\displaystyle \sum_{t=1}^\infty \|\pi_{(t)}-\pi_{(t+1)}\|_1 < \infty$.
Since we know from earlier in the proof that $\pi_{(t)}(w)$ is increasing for
$w \in \mathcal{H}$ and eventually decreasing for $w \in \mathcal{H}^C$, there
exists a $t_0 \in \mathbb{N}$ such that for any $t_1 > t_0$, we have (\ref{eq:long2}).
\begin{figure*}
\vspace*{-5mm}
\begin{equation}\label{eq:long2}
\begin{split}
\sum_{t=t_0}^{t_1} \|\pi_{(t)}-\pi_{(t+1)}\|_1=& \sum_{w \in \mathcal{H}}\sum_{t=t_0}^{t_1} \left(\pi_{(t+1)}(w)-\pi_{(t)}(w)\right) + \sum_{w \notin \mathcal{H}}\sum_{t=t_0}^{t_1} \left(\pi_{(t)}(w)-\pi_{(t+1)}(w)\right)\\
=& \sum_{w \in \mathcal{H}}\left(\pi_{(t_1+1)}(w)-\pi_{(t_0)}(w)\right) + \sum_{w \notin \mathcal{H}} \left(\pi_{(t_0)}(w)-\pi_{(t_1+1)}(w)\right)\\
=& \|\pi_{(t_1+1)}-\pi_{(t_0)}\|_1 \le \|\pi_{(t_1+1)}\|_1+\|\pi_{(t_0)}\|_1 = 2
\end{split}
\end{equation}
\end{figure*}
Since the right hand side does not depend on $t_1$, we have that
$\sum_{t=1}^\infty \|\pi_{(t)}-\pi_{(t+1)}\|_1 < \infty$. This implies that the
non-homogeneous MC used by Algorithm~\ref{alg:MCMC} is strongly ergodic, and
thus completes the proof of Theorem~\ref{th:conv}.

\section{Proof of Lemma~\ref{lemm:ergcoefbound}}
\label{app:lemm:ergcoefbound}
Let $w',w''$ be two arbitrary sequences in $\ss$. The probability of transitioning
from a given state to a neighboring state in an iteration within iteration $t'$ of
super-iteration $t$ of Algorithm~\ref{alg:MCMC} is given by (\ref{eqn:Gibbs}), and
can be rewritten as (\ref{eq:trans_prob}),
\begin{figure*}
\vspace*{-5mm}
\begin{equation}\label{eq:trans_prob}
\begin{split}
P_{(t,t')}(w_1&^{t'-1}aw_{t'+1}^N|w_1^{t'-1}bw_{t'+1}^N) = p_{s_t}(w_{t'} = a|w^{\backslash t'}) = \frac{ \exp\left(-s_t\Psi^{H_q}(w_1^{t'-1}aw_{t'+1}^N) \right) }{ \sum_{b\in\replevels_F} \exp\left( -s_t\Psi^{H_q}(w_1^{t'-1}bw_{t'+1}^N) \right) }\\
& = \frac{ \exp\left(-s_t\left(\Psi^{H_q}(w_1^{t'-1}aw_{t'+1}^N) -\Psi^{H_q}_{\min,t'}(w_1^{t'-1},w_{t'+1}^N)\right) \right) }{ \sum_{b\in \replevels_F} \exp\left( -s_t\left(\Psi^{H_q}(w_1^{t'-1}bw_{t'+1}^N)-\Psi^{H_q}_{\min,t'}(w_1^{t'-1},w_{t'+1}^N)\right) \right) } \ge \frac{\exp(-s_t\Delta_q)}{|\replevels_F|}
\end{split}
\end{equation}
\hrulefill
\end{figure*}
where $\Psi^{H_q}_{\min,t'}(w_1^{t'-1},w_{t'+1}^N) = \min_{\beta \in \replevels_F} \Psi^{H_q}(w_1^{t'-1}\beta w_{t'+1}^N)$.
Therefore, the smallest probability of transition from $w'$ to $w''$ within
super-iteration $t$ of Algorithm~\ref{alg:MCMC} is bounded by
\begin{align*}
\min_{w',w'' \in \replevels_F}P_{(t)}(w''|w') & \ge \prod_{t'=1}^{N} \frac{\exp(-s_t\Delta_q)}{|\replevels_F|}\\
&= \frac{\exp(-s_tN\Delta_q)}{|\replevels_F|^N} = \frac{\exp(-s_tN\Delta_q)}{|\ss|}.
\end{align*}
Using the alternative definition of the ergodic coefficient (\ref{eq:ergcoef}),
\begin{align*}
\xi\left(P_{(t)}\right) &= 1-\min_{w',w'' \in \ss} \sum_{\widetilde{w} \in \ss} \min(P_{(t)}(\widetilde{w}|w'),P_{(t)}(\widetilde{w}|w''))\\
&\le 1 - |\ss|\frac{\exp(-s_tN\Delta_q)}{|\ss|} = 1-\exp(-s_tN\Delta_q),
\end{align*}
proving the lemma.

\vspace*{0mm}
\section*{Acknowledgments}

Preliminary conversations with Deanna Needell and Tsachy Weissman framed our
thinking about universal compressed sensing.
Phil Schniter was instrumental in formulating the proposed framework and shepherding our progress through detailed conversations, feedback on our drafts, and
probing questions. Gary Howell provided invaluable guidance on using North Carolina State University's  high performance computing resources. Final thanks to Jin Tan, Yanting Ma, and Nikhil Krishnan for thoroughly proofreading our manuscript.


\begin{thebibliography}{10}

\bibitem{JZ2014SSP}
J.~Zhu, D.~Baron, and M.~F. Duarte,
\newblock ``Complexity--adaptive universal signal estimation for compressed
  sensing,''
\newblock in {\em Proc. IEEE Stat. Signal Process. Workshop (SSP)}, June 2014,
  pp. 416--419.

\bibitem{BaronDuarteAllerton2011}
D.~Baron and M.~F. Duarte,
\newblock ``Universal {MAP} estimation in compressed sensing,''
\newblock in {\em Proc. Allerton Conference Commun., Control, and Comput.},
  Sept. 2011, pp. 768--775.

\bibitem{BaronFinland2011}
D.~Baron,
\newblock ``Information complexity and estimation,''
\newblock in {\em Workshop Inf. Theoretic Methods Sci. Eng. (WITMSE)}, Aug.
  2011.

\bibitem{CandesRUP}
E.~Cand\`{e}s, J.~Romberg, and T.~Tao,
\newblock ``Robust uncertainty principles: {E}xact signal reconstruction from
  highly incomplete frequency information,''
\newblock {\em IEEE Trans. Inf. Theory}, vol. 52, no. 2, pp. 489--509, Feb.
  2006.

\bibitem{DonohoCS}
D.~Donoho,
\newblock ``Compressed sensing,''
\newblock {\em IEEE Trans. Inf. Theory}, vol. 52, no. 4, pp. 1289--1306, Apr.
  2006.

\bibitem{WuVerdu2012}
Y.~Wu and S.~Verd{\'u},
\newblock ``Optimal phase transitions in compressed sensing,''
\newblock {\em {IEEE} Trans. Inf. Theory}, vol. 58, no. 10, pp. 6241 -- 6263,
  Oct. 2012.

\bibitem{Donoho2013}
D.~Donoho, I.~Johnstone, and A.~Montanari,
\newblock ``Accurate prediction of phase transitions in compressed sensing via
  a connection to minimax denoising,''
\newblock {\em IEEE Trans. Inf. Theory}, vol. 59, no. 6, pp. 3396--3433, June
  2013.

\bibitem{Tan_CompressiveImage2014}
J.~Tan, Y.~Ma, and D.~Baron,
\newblock ``Compressive imaging via approximate message passing with image
  denoising,''
\newblock {\em Arxiv preprint arxiv:1405.4429}, May 2014,
\newblock submitted.

\bibitem{MaZhuBaron2014submit}
Yanting Ma, Junan Zhu, and Dror Baron,
\newblock ``Compressed sensing via universal denoising and approximate message
  passing,''
\newblock {\em Arxiv preprint arxiv:1407.1944}, July 2014,
\newblock submitted.

\bibitem{GPSR2007}
M.~Figueiredo, R.~Nowak, and S.~J. Wright,
\newblock ``Gradient projection for sparse reconstruction: Application to
  compressed sensing and other inverse problems,''
\newblock {\em IEEE J. Select. Topics Signal Proces.}, vol. 1, pp. 586--597,
  Dec. 2007.

\bibitem{DMM2010ITW1}
D.~L. Donoho, A.~Maleki, and A.~Montanari,
\newblock ``{Message Passing Algorithms for Compressed Sensing: I. Motivation
  and Construction},''
\newblock in {\em IEEE Inf. Theory Workshop}, Jan. 2010.

\bibitem{RanganGAMP2011ISIT}
S.~Rangan,
\newblock ``Generalized approximate message passing for estimation with random
  linear mixing,''
\newblock in {\em Proc. IEEE Int. Symp. Inf. Theory (ISIT)}, July 2011, pp.
  2168--2172.

\bibitem{BCS2008}
S.~Ji, Y.~Xue, and L.~Carin,
\newblock ``Bayesian compressive sensing,''
\newblock {\em {IEEE Trans. Signal Process.}}, vol. 56, no. 6, pp. 2346--2356,
  June 2008.

\bibitem{BCSEx2008}
M.~W. Seeger and H.~Nickisch,
\newblock ``Compressed sensing and {B}ayesian experimental design,''
\newblock in {\em Proc. Int. Conference Machine Learning}, Aug 2008, pp.
  912--919.

\bibitem{CSBP2010}
D.~Baron, S.~Sarvotham, and R.~G. Baraniuk,
\newblock ``Bayesian compressive sensing via belief propagation,''
\newblock {\em IEEE Trans. Signal Process.}, vol. 58, pp. 269--280, Jan. 2010.

\bibitem{EMGMTSP}
J.~Vila and P.~Schniter,
\newblock ``Expectation-maximization {G}aussian-mixture approximate message
  passing,''
\newblock {\em IEEE Trans. Signal Process.}, vol. 61, no. 19, pp. 4658--4672,
  Oct. 2013.

\bibitem{turboGAMP}
J.~Ziniel, S.~Rangan, and P.~Schniter,
\newblock ``A generalized framework for learning and recovery of structured
  sparse signals,''
\newblock in {\em Proc. IEEE Stat. Signal Process. Workshop (SSP)}, Aug. 2012,
  pp. 325--328.

\bibitem{MTKB2014ITA}
Y.~Ma, J.~Tan, N.~Krishnan, and D.~Baron,
\newblock ``Empirical {B}ayes and full {B}ayes for signal estimation,''
\newblock {\em Arxiv preprint arxiv:1405.2113v1}, May 2014.

\bibitem{Figueiredo2003}
M.~A.~T. Figueiredo and R.~D. Nowak,
\newblock ``An {EM} algorithm for wavelet-based image restoration,''
\newblock {\em IEEE Trans. Image Process.}, vol. 12, no. 8, pp. 906--916, Aug.
  2003.

\bibitem{DonohoKolmogorovCS2006}
D.~Donoho, H.~Kakavand, and J.~Mammen,
\newblock ``The simplest solution to an underdetermined system of linear
  equations,''
\newblock in {\em Proc. Int. Symp. Inf. Theory (ISIT)}, July 2006, pp.
  1924--1928.

\bibitem{HN05}
J.~Haupt and R.~Nowak,
\newblock ``{S}ignal reconstruction from noisy random projections,''
\newblock {\em IEEE Trans. Inf. Theory}, vol. 52, no. 9, pp. 4036--4048, Sept.
  2006.

\bibitem{HN11}
J.~D. Haupt and R.~Nowak,
\newblock ``Adaptive sensing for sparse recovery,''
\newblock in {\em Compressed Sensing: Theory and Applications}. Cambridge
  University Press, 2012.

\bibitem{Ramirez2011}
I.~Ram{\'\i}rez and G.~Sapiro,
\newblock ``An {MDL} framework for sparse coding and dictionary learning,''
\newblock {\em IEEE Trans. Signal Process.}, vol. 60, no. 6, pp. 2913--2927,
  June 2012.

\bibitem{Cover06}
T.~M. Cover and J.~A. Thomas,
\newblock {\em Elements of Information Theory},
\newblock New York, NY, USA: Wiley-Interscience, 2006.

\bibitem{LiVitanyi2008}
M.~Li and P.~M.~B. Vitanyi,
\newblock {\em An introduction to {K}olmogorov complexity and its
  applications},
\newblock Springer-Verlag, New York, 2008.

\bibitem{AharoEB_KSVD}
M.~Aharon, M.~Elad, and A.~Bruckstein,
\newblock ``{K-SVD}: An algorithm for designing overcomplete dictionaries for
  sparse representation,''
\newblock {\em {IEEE} Trans. Signal Process.}, vol. 54, no. 11, pp. 4311--4322,
  Nov. 2006.

\bibitem{Mairal2008}
J.~Mairal, F.~Bach, J.~Ponce, G.~Sapiro, and A.~Zisserman,
\newblock ``Supervised dictionary learning,''
\newblock in {\em Workshop Neural Inf. Process. Syst. (NIPS)}, Vancouver,
  Canada, Dec. 2008.

\bibitem{Zhoul2011}
M.~Zhou, H.~Chen, J.~Paisley, L.~Ren, L.~Li, Z.~Xing, D.~Dunson, G.~Sapiro, and
  L.~Carin,
\newblock ``Nonparametric {B}ayesian dictionary learning for analysis of noisy
  and incomplete images,''
\newblock {\em IEEE Trans. Image Process.}, vol. 21, no. 1, pp. 130--144, Jan.
  2012.

\bibitem{Garrigues07learninghorizontal}
P.~J. Garrigues and B.~A. Olshausen,
\newblock ``Learning horizontal connections in a sparse coding model of natural
  images,''
\newblock in {\em Workshop Neural Inf. Process. Syst. (NIPS)}, Dec. 2007, pp.
  1--8.

\bibitem{LZ77}
J.~Ziv and A.~Lempel,
\newblock ``{A universal algorithm for sequential data compression},''
\newblock {\em IEEE Trans. Inf. Theory}, vol. 23, no. 3, pp. 337--343, May
  1977.

\bibitem{Rissanen1983}
J.~Rissanen,
\newblock ``{A universal data compression system},''
\newblock {\em IEEE Trans. Inf. Theory}, vol. 29, no. 5, pp. 656--664, Sept.
  1983.

\bibitem{Ramirez2010}
I.~Ramirez and G.~Sapiro,
\newblock ``Universal regularizers for robust sparse coding and modeling,''
\newblock {\em IEEE Trans. Image Process.}, vol. 21, no. 9, pp. 3850--3864,
  Sept. 2012.

\bibitem{DonohoKolmogorov}
D.~L. Donoho,
\newblock ``The {K}olmogorov sampler,''
\newblock Department of Statistics Technical Report 2002-4, Stanford
  University, Stanford, CA, Jan. 2002.

\bibitem{Chaitin1966}
G.~J. Chaitin,
\newblock ``On the length of programs for computing finite binary sequences,''
\newblock {\em J. ACM}, vol. 13, no. 4, pp. 547--569, 1966.

\bibitem{Solomonoff1964}
R.~J. Solomonoff,
\newblock ``A formal theory of inductive inference. {Part I},''
\newblock {\em Inf. and Control}, vol. 7, no. 1, pp. 1--22, Mar. 1964.

\bibitem{Kolmogorov1965}
A.~N. Kolmogorov,
\newblock ``Three approaches to the quantitative definition of information,''
\newblock {\em Problems Inf. Transmission}, vol. 1, no. 1, pp. 1--7, 1965.

\bibitem{JalaliMaleki2011}
S.~Jalali and A.~Maleki,
\newblock ``Minimum complexity pursuit,''
\newblock in {\em Proc. Allerton Conference Commun., Control, Comput.}, Sept.
  2011, pp. 1764--1770.

\bibitem{JalaliMalekiRichB2014}
S.~Jalali, A.~Maleki, and R.~G. Baraniuk,
\newblock ``Minimum complexity pursuit for universal compressed sensing,''
\newblock {\em IEEE Trans. Inf. Theory}, vol. 60, no. 4, pp. 2253--2268, Apr.
  2014.

\bibitem{Rissanen1978}
J.~Rissanen,
\newblock ``Modeling by shortest data description,''
\newblock {\em Automatica}, vol. 14, no. 5, pp. 465--471, Sept. 1978.

\bibitem{schwarz1978estimating}
G.~Schwarz,
\newblock ``Estimating the dimension of a model,''
\newblock {\em Ann. Stat.}, vol. 6, no. 2, pp. 461--464, Mar. 1978.

\bibitem{Wallace1968}
C.~S. Wallace and D.~M. Boulton,
\newblock ``An information measure for classification,''
\newblock {\em Comput. J.}, vol. 11, no. 2, pp. 185--194, 1968.

\bibitem{BRY98}
A.~Barron, J.~Rissanen, and B.~Yu,
\newblock ``The minimum description length principle in coding and modeling,''
\newblock {\em IEEE Trans. Inf. Theory}, vol. 44, no. 6, pp. 2743--2760, Oct.
  1998.

\bibitem{Geman1984}
S.~Geman and D.~Geman,
\newblock ``Stochastic relaxation, {G}ibbs distributions, and the {B}ayesian
  restoration of images,''
\newblock {\em IEEE Trans. Pattern Anal. Machine Intelligence}, vol. 6, pp.
  721--741, Nov. 1984.

\bibitem{Rangan2010CISS}
S.~Rangan,
\newblock ``Estimation with random linear mixing, belief propagation and
  compressed sensing,''
\newblock in {\em Proc. IEEE 44th Conference Inf. Sci. Syst. (CISS)}, Mar.
  2010.

\bibitem{GuoWang2008}
D.~Guo and C.~C. Wang,
\newblock ``Multiuser detection of sparsely spread {CDMA},''
\newblock {\em IEEE J. Select. Areas Commun.}, vol. 26, no. 3, pp. 421--431,
  Apr. 2008.

\bibitem{Turing1950}
A.~M. Turing,
\newblock ``Computing machinery and intelligence,''
\newblock {\em Mind}, vol. 59, no. 236, pp. 433--460, Oct. 1950.

\bibitem{BaronWeissman2012}
D.~Baron and T.~Weissman,
\newblock ``An {MCMC} approach to universal lossy compression of analog
  sources,''
\newblock {\em IEEE Trans. Signal Process.}, vol. 60, pp. 5230--5240, Oct.
  2012.

\bibitem{Jalali2008}
S.~Jalali and T.~Weissman,
\newblock ``Rate-distortion via {M}arkov chain {M}onte {C}arlo,''
\newblock in {\em Proc. Int. Symp. Inf. Theory (ISIT)}, July 2008, pp.
  852--856.

\bibitem{Jalali2012}
S.~Jalali and T.~Weissman,
\newblock ``{Block and sliding-block lossy compression via MCMC},''
\newblock {\em IEEE Trans. Commun.}, vol. 60, no. 8, pp. 2187--2198, Aug. 2012.

\bibitem{Yang1997}
E.~Yang, Z.~Zhang, and T.~Berger,
\newblock ``{Fixed-slope universal lossy data compression},''
\newblock {\em IEEE Trans. Inf. Theory}, vol. 43, no. 5, pp. 1465--1476, Sept.
  1997.

\bibitem{Willems1995CTW}
F.~M.~J. Willems, Y.~M. Shtarkov, and T.~J. Tjalkens,
\newblock ``The context tree weighting method: {B}asic properties,''
\newblock {\em IEEE Trans. Inf. Theory}, vol. 41, no. 3, pp. 653--664, May
  1995.

\bibitem{Cosamp08}
D.~Needell and J.~A. Tropp,
\newblock ``Co{S}a{MP}: Iterative signal recovery from incomplete and
  inaccurate samples,''
\newblock {\em Appl. Computational Harmonic Anal.}, vol. 26, no. 3, pp.
  301--321, May 2009.

\bibitem{ZhuBaronCISS2013}
J.~Zhu and D.~Baron,
\newblock ``Performance regions in compressed sensing from noisy
  measurements,''
\newblock in {\em Proc. 2013 Conf. Inference Sci. Syst. (CISS)}, Baltimore, MD,
  Mar. 2013, pp. 1--6.

\bibitem{CandesCSdictonary2011}
E.~J. Cand\`es, Y.~C. Eldar, D.~Needell, and P.~Randall,
\newblock ``Compressed sensing with coherent and redundant dictionaries,''
\newblock {\em Appl. Computational Harmonic Anal.}, vol. 31, no. 1, pp. 59--73,
  July 2011.

\bibitem{JalaliPoor2014}
S.~Jalali and H.~V. Poor,
\newblock ``Universal compressed sensing of {M}arkov sources,''
\newblock {\em Arxiv preprint arXiv:1406.7807}, June 2014.

\bibitem{Bremaud1999}
P.~Br{\'e}maud,
\newblock {\em {Markov chains: Gibbs fields, Monte Carlo simulation, and
  queues}}, vol.~31,
\newblock Springer Verlag, 1999.

\end{thebibliography}
\end{document}